\documentclass[12pt]{article}
\usepackage{amsmath,amssymb,amsthm,booktabs}
\usepackage{mathrsfs}
\usepackage[OT1]{fontenc}
\usepackage[utf8]{inputenc}
\usepackage[colorlinks,citecolor=blue,urlcolor=blue]{hyperref}
\usepackage{txfonts}
\usepackage{indentfirst}
\usepackage{multirow}
\usepackage{float,subfig}

\usepackage{bm}
\usepackage{euscript}
\usepackage{graphicx}
\usepackage{multicol}
\usepackage[usenames,dvipsnames,svgnames,table]{xcolor}

\usepackage[round]{natbib}
\numberwithin{equation}{section} \theoremstyle{plain}
\newtheorem{theorem}{Theorem}[section]
\newtheorem{lemma}{Lemma}[section]
\newtheorem{corollary}{Corollary}[section]

\newtheorem{remark}{Remark}[section]

\numberwithin{equation}{section}

\begin{document}

\title{Multi-time state mean-variance model in continuous time
\footnotemark[1]}

\author{Shuzhen Yang
\footnotemark[2]
\footnotemark[3]}

\renewcommand{\thefootnote}{\fnsymbol{footnote}}

\footnotetext[1]{

\ \ \textbf{MSC2010 subject classification}: 91B28; 93E20; 49N10.

\ \ \textbf{OR/MS subject classification}: Finance/portfolio; dynamic programming/optimal control.}

\footnotetext[2]{Shandong University-Zhong Tai Securities Institute for Financial Studies, Shandong University, PR China, (yangsz@sdu.edu.cn).}
\footnotetext[3]{This work was supported by the National Natural Science Foundation of China (Grant No.11701330) and Young Scholars Program of Shandong University.}

\date{}
\maketitle

 \textbf{Abstract}: The objective of the continuous time mean-variance model is  to minimize the variance (risk) of an investment portfolio with a given mean  at terminal time. However, the investor can stop the investment plan at any time before the terminal time. To solve this problem, we consider minimizing  the variances  of the investment portfolio in the multi-time state.  The advantage of this multi-time state mean-variance model is that we can minimize the risk of the investment portfolio within the  investment period.  To obtain the optimal strategy of  the model, we introduce a sequence of Riccati equations, which are connected by a jump boundary condition.  Based on this equations, we establish the relationship between the means and variances in the multi-time state mean-variance model. Furthermore, we use an example to  verify that the variances of the multi-time state can affect the average of  Maximum-Drawdown of the investment portfolio.

\addcontentsline{toc}{section}{\hspace*{1.8em}Abstract}

\section{Introduction}

To balance the return (mean) and risk (variance)  in a single-period portfolio selection model,  \cite{Ma52,Ma59} proposed the mean-variance model. Since then,  many related works focused on these topics. Under some mild assumptions,  \cite{Me72} solved  the single-period problem analytically.  \cite{R89}  studied a mean-variance model in which a single stock with a constant risk-free rate is introduced. Dynamic asset allocation in a mean-variance framework was studied by  \cite{BP98}.  \cite{LN00} embedded the discrete-time multi-period mean-variance problem within a multi-objective optimization framework and obtained an optimal strategy.  By extending  the embedding technique introduced in   \cite{LN00} and applying the results from the stochastic LQ control in the continuous time case,  \cite{ZL00} investigated an optima pair for the continuous-time mean-variance problem. Further results in the  mean-variance problem include those with bankruptcy prohibition, transaction costs, and random parameters in an complete and incomplete markets (see \cite{BJPZ05, DXZ10,L04,LZ02,X05}).

The pre-committed strategies in the aforementioned multi-period and continuous time cases, differed from those of the single-period case. For further details, see   \citep{KP97}.   \cite{BC10} adopted a game theoretic approach to study the time inconsistency in the mean-variance model and  \cite{BMZ14} studied the mean-variance problem with state dependent risk aversion.

In the financial market,  for a given terminal time $T$, $Y^{\pi}(T)$ represents a portfolio asset with  strategy $\pi(\cdot)$, while $\mathbb{E}[Y^{\pi}(T)]$ and $\mathrm{Var} (Y^{\pi}(T))=\mathbb{E}\big{(}Y^{\pi}(T)-\mathbb{E}[Y^{\pi}(T)]\big{)}^2$ represent the mean and variance of $Y^{\pi}(T)$, respectively.  In the classical mean-variance model, we want to minimize the variance of the portfolio asset $\mathrm{Var} (Y^{\pi}(T))$ for a given mean  $\mathbb{E}[Y^{\pi}(T)]=L$, where $L$ is a constant. The investor can stop the investment plan at an uncertain horizon time $\tau$ before the terminal time $T$, where $\tau\leq T$.
Therefore, there are many related works on  the mean-variance portfolio model  with an uncertain horizon  time. \cite{MU06} considered static mean-variance analysis with  an uncertain time horizon. \cite{YLL08} studied the mean-variance model of a multi-period asset-liability management problem under uncertain exit time.
Furthermore, see  \citep{WLL11,YM10,Y13} for additional studies in this vein. However, in the literature of mean-variance model under
uncertain or random exit time,  we always suppose that the uncertain horizon time $\tau$ satisfies a distribution (or a conditional distribution) and investigate the related mean-variance model at time $\tau$.

However, in general, we do not know the information of $\tau$ at initial time $t_0=0$.
Given a probability space $(\Omega,\mathcal{F},P)$, notice that  for a given partition $0=t_0<t_1<\cdots<t_N=T$ of interval $[0,T]$ and $\omega\in \Omega$, there exists $ i\in \{0,1,\cdots,N-1\}$ such that $\tau(\omega)\in[t_i,t_{i+1}]$. To reduce the variance of the portfolio asset $Y^{\pi}(\cdot)$ at $\tau\in (0,T]$, we consider  minimizing the variances of the portfolio asset at multi-time state $(Y^{\pi}(t_1),Y^{\pi}(t_2),\cdots,Y^{\pi}(t_N))$ with constraint on means of   multi-time state $(Y^{\pi}(t_1),Y^{\pi}(t_2),\cdots,Y^{\pi}(t_N))$. Therefore, we introduce the following multi-time state mean-variance model:
\begin{equation}
J(\pi(\cdot))=\sum_{i=1}^N
\mathbb{E}\big{(}Y^{\pi}(t_i)-\mathbb{E}[Y^{\pi}(t_i)]\big{)}^2,\label{in-cost-1}%
\end{equation}
with constraint on the multi-time state mean,
\begin{equation}
 \mathbb{E}[Y^{\pi}(t_i)]=L_i,\ i=1,2,\cdots,N. \label{in-mean-1}
\end{equation}
In this multi-time state mean-variance model,  we can minimize the risk of the investment portfolio within the multi-time $(t_1,t_2,\cdots,t_N)$. It should be noted that the multi-time state $(Y^{\pi}(t_1),Y^{\pi}(t_2),\cdots,Y^{\pi}(t_N))$ of the investment portfolio can affect the value of each other, and we cannot solve the multi-time state mean-variance model via one classical Riccati equation directly.  To obtain the optimal strategy of  the multi-time state mean-variance model, we introduce a sequence of Riccati equations, which are connected by a jump boundary condition (see equations (\ref{ric-eq-1}) and (\ref{ric-eq-2})).  Based on this sequence of Riccati equations, we investigate  an optimal strategy (see Theorem \ref{the-1}) and establish the relationship between the means and variances of this multi-time state mean-variance model (see Lemma \ref{lem-1}).

The Maximum-Drawdown of the asset $Y^{\pi}(\cdot)$ is an important index to evaluate a strategy in the investment portfolio model, where the Maximum-Drawdown of the asset $Y^{\pi}(\cdot)$  is defined in the interval $[0,h],\ h\leq T$, by
$$
\mathrm{MD}^h_{Y^{\pi}}=\mathrm{esssup}\ \{ z  \ | \  z=Y^{\pi}(t)- Y^{\pi}(s),\ 0\leq t\leq s\leq h   \}.
$$
Based on  simulation results of the multi-time state mean-variance model (see subsection 4.2), we can see that  the constrained condition (\ref{in-mean-1}) can affect the average of  $\mathrm{MD}^h_{Y^{\pi}}$ of the portfolio asset $Y^{\pi}(\cdot)$ (see Figure \ref{fig3}). The work is most closely related to the study  of \citep{Y18}, in which the author established the necessary and sufficient conditions for stochastic differential systems with multi-time state cost functional.

The remainder of this paper is organized as follows. In Section 2, we formulate the multi-time state mean-variance model. Then, in Section 3, we investigate an optimal strategy and  establish  the relationship between multi-time state mean and variance for the proposed model. In Section 4, based on the main results of  Section 3, we compare the multi-time state mean-variance model with classical mean-variance model. Finally, we conclude  the  paper in Section 5.

\section{Multi-time state mean-variance model }

Let $W$ be a $d$-dimensional standard Brownian motion defined on a complete
filtered probability space $(\Omega,\mathcal{F},P;\{ \mathcal{F}(t)\}_{t\geq
0})$, where $\{ \mathcal{F}(t)\}_{t\geq0}$ is the $P$-augmentation of the
natural filtration generated by  $W$.  We suppose the existence of one risk-free  bond asset and $n$ risky  stock assets  that are traded in the market, where the bond satisfies the following equation:
\begin{eqnarray*}
\left\{\begin{array}{rl}
\mathrm{d}R_0(t) & \!\!\!= r(t)R_0(t)\mathrm{d}t,\;\;  t>0,\\
 R_0(0) & \!\!\!= a_0>0,
\end{array}\right.
\end{eqnarray*}
and the $i$'th ($1\leq i\leq n $) stock asset is described by
\begin{eqnarray*}
\left\{\begin{array}{rl} \mathrm{d}R_i(t) & \!\!\!=
b_i(t)R_i(t)\mathrm{d}t+ \displaystyle R_i(t) \sum_{j=1}^{{d}}\sigma_{ij}(t)\mathrm{d}W_{j}(t),\;\;
t>0,\\
 R_i(0) & \!\!\!= a_i>0,\;\;
\end{array}\right.
\end{eqnarray*}
where $r(\cdot)\in \mathbb{R}$ is the risk-free return rate of the bond, $b(\cdot)=(b_1(\cdot),\cdots,b_n(\cdot))\in{\mathbb{R}^n}$ is the expected return rate of the risky asset, and $\sigma(\cdot)=(\sigma_{1}(\cdot),\cdots,\sigma_{n}(\cdot))^{\top}\in \mathbb{R}^{n\times d} $ is the corresponding volatility matrix. Given initial capital $x>0$, $\displaystyle \gamma(\cdot)=(\gamma_1(\cdot),\cdots,\gamma_n(\cdot))\in \mathbb{R}^n$, where $\gamma_i(\cdot)=b_i(\cdot)-r(\cdot),\ 1 \leq i \leq n$. The investor's wealth $Y^{\pi}(\cdot)$ satisfies
\begin{equation}\label{asset}
\left\{\begin{array}{rl}
\!\!\!\mathrm{d}Y^{\pi}(t)  & \!\!\!=\big[r(t)Y^{\pi}(t)  +\gamma(t)\pi(t)^{\top}   \big] \mathrm{d}t+\pi(t)\sigma(t)  \mathrm{d}W(t),  \\
 \!\!\!Y^{\pi}(0) & \!\!\!=y,
\end{array}\right.
\end{equation}
where $\pi(\cdot)=(\pi_1(\cdot),\cdots,\pi_n(\cdot))\in \mathbb{R}^{n}$ is the capital invested in the risky asset $R(\cdot)=(R_1(\cdot),\cdots,R_n(\cdot))\in \mathbb{R}^n$ and  $\pi_0(\cdot)$ is the capital invested in the bond. Thus, we have
$$
\displaystyle Y^{\pi}(\cdot)=\sum_{i=0}^n\pi_i(\cdot).
$$

In this study, we consider the following multi-time state mean-variance model:
\begin{equation}
J_1(\pi(\cdot))=\sum_{i=1}^N
\mathbb{E}\big{(}Y^{\pi}(t_i)-\mathbb{E}[Y^{\pi}(t_i)]\big{)}^2,\label{cost-1}%
\end{equation}
with constraint on the multi-time state mean,
\begin{equation}
 \mathbb{E}[Y^{\pi}(t_i)]=L_i,\ i=1,2,\cdots,N, \label{mean-1}
\end{equation}
where $0=t_0<t_1<\cdots <t_N=T$.
The set of admissible strategies $\pi(\cdot)$ is defined as:
$$
\mathcal{A}=\bigg{\{}\pi(\cdot):  \pi(\cdot)\in L^2_{\mathcal{F}}[0,T;\mathbb{R}^n]\bigg{\}},
$$
where $L^2_{\mathcal{F}}[0,T;\mathbb{R}^n]$ is the set of all square integrable measurable $\mathbb{R}^n$ valued  $\{\mathcal{F}_t\}_{t\geq 0}$ adaptive processes. If there exists a  strategy $\pi^{*}(\cdot)\in \mathcal{A}$  that yields the minimum value of the cost functional (\ref{cost-1}), then we say that the multi-time state mean-variance model (\ref{cost-1}) is solved.

We make the following assumptions  to obtain the optimal strategy for the proposed  model (\ref{cost-1}):

{$\textbf{H}_1$}: $r(\cdot), b(\cdot)$ and $\sigma(\cdot)$ are bounded deterministic continuous functions.

{$\textbf{H}_2$}: $r(\cdot),\gamma(\cdot)>0$, $\sigma(\cdot)\sigma(\cdot)^{\top}>\delta  \textbf{I}$, where $\delta>0$ is a given constant and $\textbf{I}$ is the identity matrix of $\mathbb{S}^{n}$, $\mathbb{S}^{n}$ is the set of symmetric matrices.

\section{Optimal strategy}
In this section, we investigate an optimal strategy $\pi(\cdot)$ for the problem defined in (\ref{cost-1}), with a constraint on the multi-time state mean   (\ref{mean-1}).
Here,  we  describe   how to construct an optimal strategy for (\ref{cost-1}) with constrained condition (\ref{mean-1}).

Similar to \citep{ZL00}, we introduce the following multi-time state mean-variance problem: minimizing  the  cost functional,
\begin{equation}
\label{cost-3-1}
J_2(\pi(\cdot))=\sum_{i=1}^N\bigg{(}\frac{\mu_i}{2} \mathrm{Var} (Y^{\pi}(t_i))-\mathbb{E}[Y^{\pi}(t_i)]\bigg{)}.
\end{equation}
To solve the cost functional (\ref{cost-3-1}), we employ the following model:
 \begin{equation}
\label{cost-3-2}
J_3(\pi(\cdot))=\sum_{i=1}^N\mathbb{E}[\frac{\mu_i}{2} Y^{\pi}(t_i)^2-\lambda_i Y^{\pi}(t_i)],
\end{equation}
where $\mu_i>0$ and $\lambda_i\in \mathbb{R}$, $i=1,2,\cdots,N$. For  given $\mu_i>0,\ i=1,2,\cdots,N$, we suppose ${\pi}^*(\cdot)$ is an optimal strategy of cost functional (\ref{cost-3-1}). Based on Theorem 3.1 of  \citep{ZL00}, taking $\displaystyle \lambda_i=1+{\mu_i}\mathbb{E}[Y^{{\pi}^*}(t_i)],\ i=1,2,\cdots,N$, we can show that ${\pi}^*(\cdot)$ is an optimal strategy of cost functional (\ref{cost-3-2}). It should be noted that, we cannot solve the cost functional (\ref{cost-3-2}) by applying the embedding technique of \citep{ZL00} for  the multi-time state mean-variance models via the classical Riccati equation directly. This is because the value $Y^{\pi}(t_i)$ can affect $Y^{\pi}(t_{i+1})$, for $i=0,1,\cdots,N-1$.

Denoting by
\begin{equation*}
\begin{array}
[c]{rl}%
& \displaystyle \rho_i=\frac{\lambda_i}{\mu_i},\ z^{\pi}_i(t)=Y^{\pi}(t)-\rho_i,\ t\leq t_i,\ i=1,2,\cdots, N,     \\
& \beta(t)=\gamma(t)[\sigma(t)\sigma(t)^{\top}]^{-1}\gamma(t)^{\top},\ t\leq T.\\
\end{array}
\end{equation*}
Thus, the cost functional (\ref{cost-3-2}) is equivalent to
 \begin{equation}
\label{cost-3-3}
J_4(\pi(\cdot))=\sum_{i=1}^N\mathbb{E}[\frac{\mu_i}{2} z_i^{\pi}(t_i)^2],
\end{equation}
where $z_i^{\pi}(\cdot)$ satisfies
\begin{equation}\label{asset}
\left\{\begin{array}{rl}
\!\!\!\mathrm{d}z_i^{\pi}(t)  & \!\!\!=\big[r(t)z_i^{\pi}(t)  +\gamma(t)\pi(t)^{\top}   +\rho_i r(t)\big] \mathrm{d}t+\pi(t)\sigma(t)  \mathrm{d}W(t),  \\
 \!\!\!z_i^{\pi}(t_{i-1}) & \!\!\!=Y^{\pi}(t_{i-1})-\rho_i,\ t_{i-1}<t\leq t_i.
\end{array}\right.
\end{equation}

Now, we construct a new sequence of Riccati equations that are connected by a jump boundary condition, in which the jump boundary condition  can offset the interaction of $Y^{\pi}(t_{i+1})$ and  $Y^{\pi}(t_{i})$, for $i=0,1,\cdots,N-1$. We first introduce a sequence of deterministic Riccati equations:
\begin{equation}\label{ric-eq-1}
\left\{\begin{array}{rl}
\!\!\!\mathrm{d}P_i(t)  & \!\!\!=\big[ \beta(t)-2r(t)\big]P_i(t) \mathrm{d}t,  \\
 \!\!\!P_i(t_i)& \!\!\!=\displaystyle {\mu}_i+P_{i+1}(t_i),\ t_{i-1}\leq t<t_i,\ i=1,2,\cdots,N,
\end{array}\right.
\end{equation}
and related equations,
\begin{equation}\label{ric-eq-2}
\left\{\begin{array}{rl}
\!\!\!\mathrm{d}g_i(t)  & \!\!\!=\big[ (\beta(t)-r(t))g_i(t)-\rho_ir(t)P_i(t)\big] \mathrm{d}t,  \\
 \!\!\!g_i(t_i)& \!\!\!=g_{i+1}(t_i)+P_{i+1}(t_i)(\rho_{i}-\rho_{i+1}),\ t_{i-1}\leq t< t_i,\  i=1,2,\cdots,N,
\end{array}\right.
\end{equation}
where $P_{N+1}(t_N)=0,\ g_{N+1}(t_N)=0,\ \rho_{N+1}=0$. Furthermore, by a simple calculation, we can obtain,
$$
\frac{g_i(t)}{P_i(t)}=\frac{g_i(t_i)}{P_i(t_i)}e^{-\int_t^{t_i}r(s)\mathrm{d}s}
+\rho_i(1-e^{-\int_t^{t_i}r(s)\mathrm{d}s}),
\ t_{i-1}\leq t\leq t_i\ i=1,2,\cdots,N,
$$
which is used to obtain the following results.

\begin{theorem}\label{the-1}
 Let Assumptions {$\textbf{H}_1$} and {$\textbf{H}_2$} hold, there exists  an optimal strategy $\pi^*(\cdot)$ for cost functional (\ref{cost-3-3}), where the optimal strategy $\pi^*(\cdot)$ is given as follows:
\begin{equation*}
\begin{array}
[c]{rl}%
\pi^*(t)
=&\displaystyle \gamma(t)(\sigma(t)\sigma(t)^{\top})^{-1}
\big[(\rho_i-\frac{g_i(t_i)}{P_i(t_i)})e^{-\int_t^{t_i}r(s)\mathrm{d}s}-Y^{*}(t)\big], \ t_{i-1}<t\leq t_i,\\
\end{array}
\end{equation*}
where $Y^{*}(t)=z^{\pi^*}(t)+\rho_i, \ t_{i-1}<t\leq t_i$ and $i=1,2,\cdots,N$.
\end{theorem}
\noindent \textbf{Proof}: For any given $i\in \{1,2, \cdots,N\}$, $t_{i-1}<t\leq t_i$, applying It\^{o} formula to $z_i^{\pi}(t)^2P_i(t)$ and $z_i^{\pi}(t)g_i(t)$, respectively,  we have
\begin{equation*}
\begin{array}
[c]{rl}%
 &\displaystyle \frac{1}{2}\mathrm{d}z_i^{\pi}(t)^2P_i(t)\\
 =&
\displaystyle \frac{1}{2}\bigg{\{}\displaystyle 2z_i^{\pi}(t)P_i(t)\big[r(t)z_i^{\pi}(t)  +\gamma(t)\pi(t)^{\top}+\rho_ir(t)   \big]
+z_i^{\pi}(t)^2 \big[ \beta(t)-2r(t)\big]P_i(t)\\
&+ P_i(t) \pi(t)\sigma(t)\sigma(t)^{\top}  \pi(t)^{\top} \bigg{\}}\mathrm{d}t+z_i^{\pi}(t)P_i(t)\pi(t)\sigma(t)\mathrm{d}W(t)\\
=&\displaystyle \frac{1}{2}\bigg{\{}\displaystyle 2z_i^{\pi}(t)P_i(t)\big[\gamma(t)\pi(t)^{\top}+\rho_ir(t)   \big]
+z_i^{\pi}(t)^2\beta(t)P_i(t)\\
&+ P_i(t) \pi(t)\sigma(t)\sigma(t)^{\top}  \pi(t)^{\top} \bigg{\}}\mathrm{d}t+z_i^{\pi}(t)P_i(t)\pi(t)\sigma(t)\mathrm{d}W(t)\\
\end{array}
\end{equation*}
and
\begin{equation*}
\begin{array}
[c]{rl}%
 &\displaystyle \mathrm{d}z_i^{\pi}(t)g_i(t)\\
 =&
\bigg{\{}\displaystyle g_i(t)\gamma(t)\pi(t)^{\top} +g_i(t)\rho_ir(t)
+z_i^{\pi}(t)\big[\beta(t)g_i(t)-\rho_ir(t)P_i(t)\big]  \bigg{\}}\mathrm{d}t\\
&+g_i(t)\pi(t)\sigma(t)\mathrm{d}W(t).\\
\end{array}
\end{equation*}
We add the above  two equations together  and integrate from $t_{i-1}$ to $t_i$, it follows that
\begin{equation*}
\begin{array}
[c]{rl}%
 &\displaystyle \mathbb{E}\bigg[\frac{P_{i}(t_i)}{2} z_i^{\pi}(t_i)^2
 -\frac{P_{i}(t_{i-1})}{2} z_{i}^{\pi}(t_{i-1})^2+z_i^{\pi}(t_i)g_i(t_i)-z_{i}^{\pi}(t_{i-1})g_{i}(t_{i-1})\bigg]\\
= &\displaystyle \mathbb{E}\bigg[\frac{\mu_i+P_{i+1}(t_i)}{2} z_i^{\pi}(t_i)^2
 -\frac{P_{i}(t_{i-1})}{2} z_{i}^{\pi}(t_{i-1})^2\\
 &+
 z_i^{\pi}(t_i)\big[g_{i+1}(t_i)
 +P_{i+1}(t_i)(\rho_{i}-\rho_{i+1})\big]-z_{i}^{\pi}(t_{i-1})g_{i}(t_{i-1})\bigg]\\
 =&  \displaystyle \mathbb{E}\bigg[\frac{\mu_i+P_{i+1}(t_i)}{2} z_i^{\pi}(t_i)^2
 -\frac{P_{i}(t_{i-1})}{2} \big[z_{i-1}^{\pi}(t_{i-1})+\rho_{i-1}-\rho_{i}\big]^2\\
& +z_i^{\pi}(t_i)\big[g_{i+1}(t_i)+P_{i+1}(t_i)(\rho_{i}-\rho_{i+1})\big]
-\big[z_{i-1}^{\pi}(t_{i-1})+\rho_{i-1}-\rho_{i}\big]g_{i}(t_{i-1})\bigg]          \\
 =&  \displaystyle \mathbb{E}\bigg[\frac{\mu_i}{2} z_i^{\pi}(t_i)^2
 -\big(\rho_{i-1}-\rho_{i}\big)^2\frac{P_{i}(t_{i-1})}{2}
 - (\rho_{i-1}-\rho_{i}) g_{i}(t_{i-1}) \\
&\displaystyle +\frac{P_{i+1}(t_i)}{2} z_i^{\pi}(t_i)^2
+P_{i+1}(t_i)(\rho_{i}-\rho_{i+1})z_i^{\pi}(t_i)
+z_i^{\pi}(t_i)g_{i+1}(t_i)
\\
& \displaystyle -\frac{P_{i}(t_{i-1})}{2} z_{i-1}^{\pi}(t_{i-1})^2
-{P_{i}(t_{i-1})}(\rho_{i-1}-\rho_{i})z_{i-1}^{\pi}(t_{i-1})
-z_{i-1}^{\pi}(t_{i-1})g_{i}(t_{i-1}) \bigg]
     \\
=&\displaystyle \frac{1}{2} \mathbb{E}\int_{t_{i-1}}^{t_i}\bigg{\{}P_i(t) \pi(t)\sigma(t)\sigma(t)^{\top}  \pi(t)^{\top}+
 2\gamma(t)\pi(t)^{\top}(z_i^{\pi}(t)P_i(t)+g_i(t)) \\
 & + z_i^{\pi}(t)^2\beta(t)P_i(t)+2z_i^{\pi}(t)\beta(t)g_i(t)+2g_i(t)\rho_ir(t) \bigg{\}}\mathrm{d}t\\
 =&\displaystyle \frac{1}{2} \mathbb{E}\int_{t_{i-1}}^{t_i}\bigg{\{}
 \big[ \pi(t) +\gamma(t)(\sigma(t)\sigma(t)^{\top})^{-1}(z_i^{\pi}(t)+\frac{g_i(t)}{P_i(t)}) \big]\sigma(t)P_i(t)\sigma(t)^{\top}\\
 &\displaystyle \big[ \pi(t) +\gamma(t)(\sigma(t)\sigma(t)^{\top})^{-1}(z_i^{\pi}(t)+\frac{g_i(t)}{P_i(t)}) \big]^{\top}\\
 &-\gamma(t)(P_i(t)\sigma(t)\sigma(t)^{\top})^{-1}\gamma(t)^{\top}g_i(t)^2 +2g_i(t)\rho_ir(t)
  \bigg{\}}\mathrm{d}t,\\
\end{array}
\end{equation*}
the third equality is derived by the following results,
\begin{equation*}
\begin{array}
[c]{rl}%
z_{i}^{\pi}(t_{i-1})&=Y^{\pi}(t_{i-1})-\rho_{i}\\
&=Y^{\pi}(t_{i-1})-\rho_{i-1}+\rho_{i-1}-\rho_{i}\\
&=z_{i-1}^{\pi}(t_{i-1})+\rho_{i-1}-\rho_{i},
\end{array}
\end{equation*}
where $z_{0}^{\pi}(t_{0})=y,\ \rho_{0}=0.$

Thus, we have
\begin{equation}\label{var-4}
\begin{array}
[c]{rl}%
&  \displaystyle \mathbb{E}\bigg[\frac{\mu_i}{2} z_i^{\pi}(t_i)^2
 -\big(\rho_{i-1}-\rho_{i}\big)^2\frac{P_{i}(t_{i-1})}{2}
 - (\rho_{i-1}-\rho_{i}) g_{i}(t_{i-1}) \\
&\displaystyle +\frac{P_{i+1}(t_i)}{2} z_i^{\pi}(t_i)^2
+P_{i+1}(t_i)(\rho_{i}-\rho_{i+1})z_i^{\pi}(t_i)
+z_i^{\pi}(t_i)g_{i+1}(t_i)
\\
& \displaystyle -\frac{P_{i}(t_{i-1})}{2} z_{i-1}^{\pi}(t_{i-1})^2
-{P_{i}(t_{i-1})}(\rho_{i-1}-\rho_{i})z_{i-1}^{\pi}(t_{i-1})
-z_{i-1}^{\pi}(t_{i-1})g_{i}(t_{i-1}) \bigg]
     \\
 =&\displaystyle \frac{1}{2} \mathbb{E}\int_{t_{i-1}}^{t_i}\bigg{\{}
 \big[ \pi(t) +\gamma(t)(\sigma(t)\sigma(t)^{\top})^{-1}(z_i^{\pi}(t)+\frac{g_i(t)}{P_i(t)}) \big]\sigma(t)P_i(t)\sigma(t)^{\top}\\
 &\displaystyle \big[ \pi(t) +\gamma(t)(\sigma(t)\sigma(t)^{\top})^{-1}(z_i^{\pi}(t)+\frac{g_i(t)}{P_i(t)}) \big]^{\top}\\
 &-\gamma(t)(P_i(t)\sigma(t)\sigma(t)^{\top})^{-1}\gamma(t)^{\top}g_i(t)^2 +2g_i(t)\rho_ir(t)
  \bigg{\}}\mathrm{d}t,\\
\end{array}
\end{equation}
Adding $i$ on both sides of equation (\ref{var-4}) from $1$ to $N$, it follows that
\begin{equation}\label{var-5}
\begin{array}
[c]{rl}%
&  \displaystyle \sum_{i=1}^N\mathbb{E}\bigg[\frac{\mu_i}{2} z_i^{\pi}(t_i)^2
 -\big(\rho_{i-1}-\rho_{i}\big)^2\frac{P_{i}(t_{i-1})}{2}
 - (\rho_{i-1}-\rho_{i}) g_{i}(t_{i-1}) \\
&\displaystyle +\frac{P_{i+1}(t_i)}{2} z_i^{\pi}(t_i)^2
+P_{i+1}(t_i)(\rho_{i}-\rho_{i+1})z_i^{\pi}(t_i)
+z_i^{\pi}(t_i)g_{i+1}(t_i)
\\
& \displaystyle -\frac{P_{i}(t_{i-1})}{2} z_{i-1}^{\pi}(t_{i-1})^2
-{P_{i}(t_{i-1})}(\rho_{i-1}-\rho_{i})z_{i-1}^{\pi}(t_{i-1})
-z_{i-1}^{\pi}(t_{i-1})g_{i}(t_{i-1}) \bigg]
     \\
=&  \displaystyle \sum_{i=1}^N\mathbb{E}\bigg[\frac{\mu_i}{2} z_i^{\pi}(t_i)^2
 -\big(\rho_{i-1}-\rho_{i}\big)^2\frac{P_{i}(t_{i-1})}{2}
 - (\rho_{i-1}-\rho_{i}) g_{i}(t_{i-1}) \bigg]\\
&\displaystyle -\mathbb{E}\bigg[ \frac{P_{1}(t_{0})}{2} z_{0}^{\pi}(t_{0})^2
+{P_{1}(t_{0})}(\rho_{0}-\rho_{1})z_{0}^{\pi}(t_{0})
+z_{0}^{\pi}(t_{0})g_{1}(t_{0}) \bigg]
     \\
 =&\displaystyle \sum_{i=1}^N\frac{1}{2} \mathbb{E}\int_{t_{i-1}}^{t_i}\bigg{\{}
 \big[ \pi(t) +\gamma(t)(\sigma(t)\sigma(t)^{\top})^{-1}(z_i^{\pi}(t)+\frac{g_i(t)}{P_i(t)}) \big]\sigma(t)P_i(t)\sigma(t)^{\top}\\
 &\displaystyle \big[ \pi(t) +\gamma(t)(\sigma(t)\sigma(t)^{\top})^{-1}(z_i^{\pi}(t)+\frac{g_i(t)}{P_i(t)}) \big]^{\top}\\
 &-\gamma(t)(P_i(t)\sigma(t)\sigma(t)^{\top})^{-1}\gamma(t)^{\top}g_i(t)^2 +2g_i(t)\rho_ir(t)
  \bigg{\}}\mathrm{d}t,\\
\end{array}
\end{equation}
and thus
\begin{equation}\label{var-6}
\begin{array}
[c]{rl}%
&  \displaystyle \mathbb{E}\bigg[\sum_{i=1}^N\frac{\mu_i}{2} z_i^{\pi}(t_i)^2 \bigg]     \\
 =&\displaystyle \sum_{i=1}^N\frac{1}{2} \mathbb{E}\int_{t_{i-1}}^{t_i}\bigg{\{}
 \big[ \pi(t)
 +\gamma(t)(\sigma(t)\sigma(t)^{\top})^{-1}(z_i^{\pi}(t)+\frac{g_i(t)}{P_i(t)}) \big]\sigma(t)\sigma(t)^{\top}\\
 &\displaystyle \big[ \pi(t) +\gamma(t)(\sigma(t)\sigma(t)^{\top})^{-1}(z_i^{\pi}(t)+\frac{g_i(t)}{P_i(t)}) \big]^{\top}\\
 &-\gamma(t)(P_i(t)\sigma(t)\sigma(t)^{\top})^{-1}\gamma(t)^{\top}g_i(t)^2 +2g_i(t)\rho_ir(t)
  \bigg{\}}\mathrm{d}t\\
  &+  \displaystyle \sum_{i=1}^N\mathbb{E}\bigg[
 \big(\rho_{i-1}-\rho_{i}\big)^2\frac{P_{i}(t_{i-1})}{2}
 + (\rho_{i-1}-\rho_{i}) g_{i}(t_{i-1}) \bigg]\\
&\displaystyle +\mathbb{E}\bigg[ \frac{P_{1}(t_{0})}{2} z_{0}^{\pi}(t_{0})^2
+{P_{1}(t_{0})}(\rho_{0}-\rho_{1})z_{0}^{\pi}(t_{0})
+z_{0}^{\pi}(t_{0})g_{1}(t_{0}) \bigg].     \\
\end{array}
\end{equation}

Based on the representation of $\displaystyle \mathbb{E}\big[\sum_{i=1}^N\frac{\mu_i}{2} z_i^{\pi}(t_i)^2 \big]$, we can obtain an optimal strategy $\pi^*(\cdot)$ for $J_4(\pi(\cdot))$, for $t\in (t_{i-1},t_i],\ i=1,2,\cdots,N$,
\begin{equation*}
\begin{array}
[c]{rl}%
\pi^*(t)=\displaystyle -\gamma(t)(\sigma(t)\sigma(t)^{\top})^{-1}(z_i^{\pi^*}(t)+\frac{g_i(t)}{P_i(t)}).
 \end{array}
\end{equation*}
Note that
$$
\frac{g_i(t)}{P_i(t)}=\frac{g_i(t_i)}{P_i(t_i)}e^{-\int_t^{t_i}r(s)\mathrm{d}s}
+\rho_i(1-e^{-\int_t^{t_i}r(s)\mathrm{d}s}),
\ t_{i-1}< t\leq t_i,
$$
where
$$
\frac{g_i(t_i)}{P_i(t_i)}=\frac
{g_{i+1}(t_i)+P_{i+1}(t_i)(\rho_{i}-\rho_{i+1})}{\mu_i+P_{i+1}(t_i)}, \ i=1,2,\cdots,N,
$$
which leads to
\begin{equation*}
\begin{array}
[c]{rl}%
\pi^*(t)
=&\displaystyle \gamma(t)(\sigma(t)\sigma(t)^{\top})^{-1}
\big[\big(\rho_i-\frac{g_i(t_i)}{P_i(t_i)}\big)e^{-\int_t^{t_i}r(s)\mathrm{d}s}-Y^{*}(t)\big], \ t_{i-1}<t\leq t_i,\\
 \end{array}
\end{equation*}
where $Y^{*}(t)=z^{\pi^*}(t)+\rho_i, \ t_{i-1}<t\leq t_i$ and $i=1,2,\cdots,N$.

\noindent This completes the proof. $\ \ \ \ \ \ \ \ \ \ \ \Box$

\bigskip

Now, we consider the process of portfolio asset equation according to $\pi^*(\cdot)$,
\begin{equation}
\left\{\begin{array}{rl}
\mathrm{d}Y^{*}(t)  &=\big[r(t)Y^{*}(t)  +\gamma(t)\pi^*(t)^{\top}   \big] \mathrm{d}t+\pi^*(t)\sigma(t)  \mathrm{d}W(t),  \\
Y^{*}(0) &=y.
\end{array}\right.
\end{equation}
$\mathbb{E}[{Y}^{*}(\cdot)]$ and $\mathbb{E}[{Y}^{*}(\cdot)^2]$ satisfy the following linear ordinary differential equations:
\begin{equation}\label{step-asset-1}
\left\{\begin{array}{rl}
\mathrm{d}\mathbb{E}[Y^{*}(t)]  & =\bigg[(r(t)-\beta(t))\mathbb{E}[{Y}^{*}(t)]
 \displaystyle+\big(\rho_i-\frac{g_i(t_i)}{P_i(t_i)}\big)e^{-\int_t^{t_i}r(s)\mathrm{d}s}\beta(t)  \bigg] \mathrm{d}t,  \\
Y^{*}(0) & =y, \ t_{i-1}< t\leq t_i, \ i=1,2,\cdots, N,
\end{array}\right.
\end{equation}
and
\begin{equation}\label{step-asset-2}
\left\{\begin{array}{rl}
\mathrm{d}\mathbb{E}[Y^{*}(t)^2]  & =\bigg[(2r(t)-\beta(t))\mathbb{E}[{Y}^{*}(t)^2]
\displaystyle +\big(\rho_i-\frac{g_i(t_i)}{P_i(t_i)}\big)^2e^{-\int_t^{t_i}2r(s)\mathrm{d}s}\beta(t)  \bigg] \mathrm{d}t,  \\
Y^{*}(0)^2 & =y^2, \ t_{i-1}< t\leq t_i, \ i=1,2,\cdots, N.
\end{array}\right.
\end{equation}

In the following, we investigate the efficient frontier of the multi-time state mean-variance  $\mathrm{Var}(Y^{*}(t_i))$ and $\mathbb{E}[Y^{*}(t_i)]$.
\begin{lemma}\label{lem-1}
Let Assumptions {$\textbf{H}_1$} and {$\textbf{H}_2$} hold, the relationship of  $\mathrm{Var}(Y^{*}(t_i))$ and $\mathbb{E}[Y^{*}(t_i)]$ is given as follows:
\begin{equation}
\begin{array}
[c]{rl}%
&\mathrm{Var}(Y^{*}(t_i))
=\displaystyle \mathrm{Var}(Y^{*}(t_{i-1}))
e^{\int_{t_{i-1}}^{t_i}[2r(t)-\beta(t)]\mathrm{d}t}+\frac{\bigg(\mathbb{E}[Y^{*}(t_i)]
-\mathbb{E}[Y^{*}(t_{i-1})]e^{\int_{t_{i-1}}^{t_i}r(t)\mathrm{d}t}\bigg)^2}
{e^{\int_{t_{i-1}}^{t_i}\beta(t)\mathrm{d}t}-1},\\
 \end{array}
\end{equation}
where $i=1,2,\cdots,N$.
\end{lemma}
\noindent \textbf{Proof}: Combining equations (\ref{step-asset-1}) and (\ref{step-asset-2}), we have for $i=1,2,\cdots, N,$
\begin{equation}\label{mean-2}
\mathbb{E}[Y^{*}(t_i)]=
\mathbb{E}[Y^{*}(t_{i-1})]e^{\int_{t_{i-1}}^{t_i}[r(t)-\beta(t)]\mathrm{d}t}
+\displaystyle\big(\rho_i-\frac{g_i(t_i)}{P_i(t_i)}\big)
\big(1-e^{-\int_{t_{i-1}}^{t_i}\beta(t)\mathrm{d}t}\big),
\end{equation}
and
\begin{equation}\label{var-2}
\mathbb{E}[Y^{*}(t_i)^2]=
\mathbb{E}[Y^{*}(t_{i-1})^2]e^{\int_{t_{i-1}}^{t_i}[2r(t)-\beta(t)]\mathrm{d}t}
+\displaystyle\big(\rho_i-\frac{g_i(t_i)}{P_i(t_i)}\big)^2
\big(1-e^{-\int_{t_{i-1}}^{t_i}\beta(t)\mathrm{d}t}\big).
\end{equation}
By equation (\ref{mean-2}), we have
$$
\displaystyle\rho_i-\frac{g_i(t_i)}{P_i(t_i)}=
\frac{\mathbb{E}[Y^{*}(t_i)]-\mathbb{E}[Y^{*}(t_{i-1})]e^{\int_{t_{i-1}}^{t_i}[r(t)
-\beta(t)]\mathrm{d}t}}
{1-e^{-\int_{t_{i-1}}^{t_i}\beta(t)\mathrm{d}t}}.
$$
Plugging $\displaystyle\rho_i-\frac{g_i(t_i)}{P_i(t_i)}$ into equation (\ref{var-2}), it follows that
\begin{equation*}
\begin{array}
[c]{rl}%
\mathbb{E}[Y^{*}(t_i)^2]=
\mathbb{E}[Y^{*}(t_{i-1})^2]e^{\int_{t_{i-1}}^{t_i}[2r(t)-\beta(t)]\mathrm{d}t}
+\displaystyle
\frac{\bigg(\mathbb{E}[Y^{*}(t_i)]-\mathbb{E}[Y^{*}(t_{i-1})]e^{\int_{t_{i-1}}^{t_i}[r(t)
-\beta(t)]\mathrm{d}t}\bigg)^2}
{1-e^{-\int_{t_{i-1}}^{t_i}\beta(t)\mathrm{d}t}},
 \end{array}
\end{equation*}
and thus
\begin{equation*}
\begin{array}
[c]{rl}%
&\mathrm{Var}(Y^{*}(t_i))(1-e^{-\int_{t_{i-1}}^{t_i}\beta(t)\mathrm{d}t})\\
=
&\big(\mathbb{E}[Y^{*}(t_{i-1})^2]-
\big[\mathbb{E}Y^{*}(t_{i-1})\big]^2\big)e^{\int_{t_{i-1}}^{t_i}[2r(t)-\beta(t)]\mathrm{d}t}
(1-e^{-\int_{t_{i-1}}^{t_i}\beta(t)\mathrm{d}t})\\
&+[\mathbb{E}Y^{*}(t_{i-1})]^2\big(e^{\int_{t_{i-1}}^{t_i}[2r(t)-\beta(t)]\mathrm{d}t}
-e^{\int_{t_{i-1}}^{t_i}[2r(t)-2\beta(t)]\mathrm{d}t}\big)\\
&+\displaystyle
{\bigg(\mathbb{E}[Y^{*}(t_i)]-\mathbb{E}[Y^{*}(t_{i-1})]e^{\int_{t_{i-1}}^{t_i}[r(t)
-\beta(t)]\mathrm{d}t}\bigg)^2}
+(e^{-\int_{t_{i-1}}^{t_i}\beta(s)\mathrm{d}s}-1)\big[\mathbb{E}Y^{*}(t_{i})\big]^2\\
=&\big(\mathbb{E}[Y^{*}(t_{i-1})^2]-
\big[\mathbb{E}Y^{*}(t_{i-1})\big]^2\big)e^{\int_{t_{i-1}}^{t_i}[2r(t)-\beta(t)]\mathrm{d}t}
(1-e^{-\int_{t_{i-1}}^{t_i}\beta(t)\mathrm{d}t})\\
&+[\mathbb{E}Y^{*}(t_{i-1})]^2e^{\int_{t_{i-1}}^{t_i}[2r(t)-\beta(t)]\mathrm{d}t}
+\displaystyle
[\mathbb{E}Y^{*}(t_i)]^2e^{-\int_{t_{i-1}}^{t_i}\beta(t)\mathrm{d}t}\\
&-2\mathbb{E}[Y^{*}(t_i)]
\mathbb{E}[Y^{*}(t_{i-1})]e^{\int_{t_{i-1}}^{t_i}[r(t)
-\beta(t)]\mathrm{d}t}
\\
=&\big(\mathbb{E}[Y^{*}(t_{i-1})^2]-
\big[\mathbb{E}Y^{*}(t_{i-1})\big]^2\big)e^{\int_{t_{i-1}}^{t_i}[2r(t)-\beta(t)]\mathrm{d}t}
(1-e^{-\int_{t_{i-1}}^{t_i}\beta(t)\mathrm{d}t})\\
&+e^{-\int_{t_{i-1}}^{t_i}\beta(t)\mathrm{d}t}\bigg(\mathbb{E}[Y^{*}(t_i)]
-\mathbb{E}[Y^{*}(t_{i-1})]e^{\int_{t_{i-1}}^{t_i}r(t)\mathrm{d}t}\bigg)^2,\\
 \end{array}
\end{equation*}
which deduces that
\begin{equation*}
\begin{array}
[c]{rl}%
\mathrm{Var}(Y^{*}(t_i))=&\mathrm{Var}(Y^{*}(t_{i-1}))
e^{\int_{t_{i-1}}^{t_i}[2r(t)-\beta(t)]\mathrm{d}t}
+\displaystyle\frac{\bigg(\mathbb{E}[Y^{*}(t_i)]
-\mathbb{E}[Y^{*}(t_{i-1})]e^{\int_{t_{i-1}}^{t_i}r(t)\mathrm{d}t}\bigg)^2}
{e^{\int_{t_{i-1}}^{t_i}\beta(t)\mathrm{d}t}-1}.\\
 \end{array}
\end{equation*}
\noindent This completes the proof. $\ \ \ \ \ \ \ \ \ \ \ \Box$
\begin{remark}
Specially, for $i=1$, one obtains
\begin{equation*}
\begin{array}
[c]{rl}%
\mathrm{Var}(Y^{*}(t_1))=&\displaystyle\frac{\bigg(\mathbb{E}[Y^{*}(t_1)]
-ye^{\int_{t_{0}}^{t_1}r(t)\mathrm{d}t}\bigg)^2}
{e^{\int_{t_{0}}^{t_1}\beta(t)\mathrm{d}t}-1},\\
 \end{array}
\end{equation*}
which is the same as the efficient frontier in  \citep{ZL00}.
\end{remark}

It should be noted that the optimal strategy $\pi^*(\cdot)$ of cost functional (\ref{cost-3-3}) depends on the parameters $\mu=(\mu_1,\cdots,\mu_N),\ \lambda=(\lambda_1,\cdots,\lambda_N)\in \mathbb{R}^N$. We want to show that there exist $\lambda$ and $\mu$ such that  the optimal strategy $\pi^*(\cdot)$ of cost functional (\ref{cost-3-3}) is an optimal strategy of cost functional (\ref{cost-3-2}).
\begin{theorem}\label{the-3}
Let Assumptions {$\textbf{H}_1$}, {$\textbf{H}_2$} hold, and
\begin{equation}\label{con-1}
\begin{array}
[c]{rl}%
&L_i
-L_{i-1}e^{\int_{t_{i-1}}^{t_{i}}r(t)\mathrm{d}t}>0,\ \ i=1,2,\cdots,N;\\
 &\displaystyle { \big[1+P_{i+1}(t_{i})\rho_{i+1}-g_{i+1}(t_{i}) \big]
(1-e^{-\int_{t_{i-1}}^{t_{i}}\beta(t)\mathrm{d}t})}\\
> & {\big[L_{i}-L_{i-1}
e^{\int_{t_{i-1}}^{t_{i}}[r(t)-\beta(t)]\mathrm{d}t}\big]P_{i+1}(t_{i})}, \ \ i=1,2,\cdots,N-1,
\end{array}
\end{equation}
where  $L_0=y$. There exists $\lambda^*=(\lambda_1^*,\lambda_2^*,\cdots,\lambda_N^*),\ \mu=(\mu_1,\mu_2,\cdots,\mu_N)\in \mathbb{R}^N$ which are determined by
 \begin{equation}
\lambda^*_i=1+{\mu_i}\mathbb{E}[Y^{*}(t_i)],\ \rho_i=\frac{\lambda^*_i}{\mu_i},\ i=1,2,\cdots,N,
\end{equation}
 such that the optimal strategy $\pi^*(\cdot)$ of cost functional (\ref{cost-3-3}) is an optimal strategy of cost functional (\ref{cost-3-2}).
\end{theorem}
\noindent \textbf{Proof}: By Theorem \ref{the-1}, an optimal strategy of model (\ref{cost-3-2}) can be solved by
(\ref{cost-3-3}), let
\begin{equation}\label{mean-3}
\lambda^*_i=1+{\mu_i}\mathbb{E}[Y^{*}(t_i)],\ \rho_i=\frac{\lambda^*_i}{\mu_i},\ i=1,2,\cdots,N.
\end{equation}
Note that $\mathbb{E}[Y^{*}(t_i)]$ depends on $\lambda^*_i$. To solve the parameters $\lambda^*_i, i=1,2,\cdots, N$, by equation (\ref{mean-2}), we first consider the case $i=N$,
\begin{equation}\label{equa-1}
\mathbb{E}[Y^{*}(t_N)]=
\mathbb{E}[Y^{*}(t_{N-1})]e^{\int_{t_{N-1}}^{t_N}[r(t)-\beta(t)]\mathrm{d}t}
+
\frac{\lambda^*_N}{\mu_N}\big[1-e^{-\int_{t_{N-1}}^{t_N}\beta(t)\mathrm{d}t}\big]
\end{equation}
and
$$
\lambda^*_N=1+\mu_N\mathbb{E}[Y^{*}(t_{N-1})]e^{\int_{t_{N-1}}^{t_N}[r(t)-\beta(t)]\mathrm{d}t}
+{\lambda^*_N}\big[1-e^{-\int_{t_{N-1}}^{t_N}\beta(t)\mathrm{d}t}\big].
$$
Thus, we have
\begin{equation}\label{lam-1}
\lambda^*_N=
{e^{\int_{t_{N-1}}^{t_N}\beta(t)\mathrm{d}t}
+\mu_N\mathbb{E}[Y^{*}(t_{N-1})]e^{\int_{t_{N-1}}^{t_N}r(t)\mathrm{d}t}}.
\end{equation}
Based on the representation of $\lambda^*_N$, by equation (\ref{equa-1}), we have
$$
\mathbb{E}[Y^*(t_N)]=\frac{e^{\int_{t_{N-1}}^{t_N}\beta(t)\mathrm{d}t}-1}
{\mu_N}+{\mathbb{E}[Y^*(t_{N-1})]e^{\int_{t_{N-1}}^{t_N}r(t)\mathrm{d}t}}
,
$$
which indicates that
\begin{equation}\label{mu-1}
\mu_N=\frac{e^{\int_{t_{N-1}}^{t_N}\beta(t)\mathrm{d}t}-1}
{\mathbb{E}[Y^*(t_N)]-
\mathbb{E}[Y^*(t_{N-1})]e^{\int_{t_{N-1}}^{t_N}r(t)\mathrm{d}t}}.
\end{equation}
Based on constrained condition (\ref{mean-1}) of $\mathbb{E}[Y^*(t_N)]=L_N,\ \mathbb{E}[Y^*(t_{N-1})]=L_{N-1}$ and condition (\ref{con-1}), we can solve $\lambda^*_N$ and $\mu_N>0$.

In the following, we consider the case $i=N-1$. By equations (\ref{mean-2}) and (\ref{mean-3}), one obtains
$$
\lambda^*_{N-1}=1+{\mu_{N-1}}\mathbb{E}[Y^{*}(t_{N-1})],
$$
\begin{equation*}
\begin{array}
[c]{rl}%
&\mathbb{E}[Y^{*}(t_{N-1})]
-\mathbb{E}[Y^{*}(t_{N-2})]e^{\int_{t_{N-2}}^{t_{N-1}}[r(t)-\beta(t)]\mathrm{d}t}\\
=
&\displaystyle\big(\rho_{N-1}-\frac{g_{N-1}(t_{N-1})}{P_{N-1}(t_{N-1})}\big)
\big[1-e^{-\int_{t_{N-2}}^{t_{N-1}}\beta(t)\mathrm{d}t}\big]\\
=&\displaystyle\big(\frac{\lambda^*_{N-1}}{\mu_{N-1}}
-\frac
{g_{N}(t_{N-1})+P_{N}(t_{N-1})(\rho_{{N-1}}-\rho_{N})}{\mu_{N-1}+P_{N}(t_{N-1})}\big)
\big[1-e^{-\int_{t_{N-2}}^{t_{N-1}}\beta(t)\mathrm{d}t}\big]\\
=&\displaystyle
\frac{\lambda^*_{N-1}+P_{N}(t_{N-1})\rho_{N}-g_{N}(t_{N-1})}{\mu_{N-1}+P_{N}(t_{N-1})}
\big[1-e^{-\int_{t_{N-2}}^{t_{N-1}}\beta(t)\mathrm{d}t}\big],
 \end{array}
\end{equation*}
and thus
\begin{equation}\label{ee-2}
\begin{array}
[c]{rl}%
&\mathbb{E}[Y^{*}(t_{N-1})]
-\displaystyle
\mathbb{E}[Y^{*}(t_{N-2})]e^{\int_{t_{N-2}}^{t_{N-1}}[r(t)-\beta(t)]\mathrm{d}t}\\
&=\displaystyle
\frac{\lambda^*_{N-1}+P_{N}(t_{N-1})\rho_{N}-g_{N}(t_{N-1})}{\mu_{N-1}+P_{N}(t_{N-1})}
\big[1-e^{-\int_{t_{N-2}}^{t_{N-1}}\beta(t)\mathrm{d}t}\big].
 \end{array}
\end{equation}
It follows that,
\begin{equation*}
\begin{array}
[c]{rl}%
\lambda^*_{N-1}=&1+\mu_{N-1}\mathbb{E}[Y^{*}(t_{N-1})]\\
=&\displaystyle
1+\frac{P_{N}(t_{N-1})\rho_{N}-g_{N}(t_{N-1})}{u_{N-1}+P_{N}(t_{N-1})}
\big[1-e^{-\int_{t_{N-2}}^{t_{N-1}}\beta(t)\mathrm{d}t}\big]\mu_{N-1}\\
&+\displaystyle\mu_{N-1}\mathbb{E}[Y^{*}(t_{N-2})]e^{\int_{t_{N-2}}^{t_{N-1}}[r(t)-\beta(t)]\mathrm{d}t}
+
\frac{\mu_{N-1}\lambda^*_{N-1}}{u_{N-1}+P_{N}(t_{N-1})}
\big[1-e^{-\int_{t_{N-2}}^{t_{N-1}}\beta(t)\mathrm{d}t}\big].
 \end{array}
\end{equation*}
Note that the coefficient of $\lambda^*_{N-1}$ is
$$
I^*_{N-1}=\frac{P_{N}(t_{N-1})
+\mu_{N-1}e^{-\int_{t_{N-2}}^{t_{N-1}}\beta(t)\mathrm{d}t}}{\mu_{N-1}+P_{N}(t_{N-1})}
>0,
$$
which indicates that there exists a unique solution for $\lambda^*_{N-1}$:
\begin{equation}\label{mean-4}
\begin{array}
[c]{rl}%
\lambda^*_{N-1}=& \displaystyle   \frac{P_N(t_{N-1})+\mu_{N-1}}{P_{N}(t_{N-1})
+\mu_{N-1}e^{-\int_{t_{N-2}}^{t_{N-1}}\beta(t)\mathrm{d}t}}\\
&\displaystyle +\mu_{N-1}\bigg{(}
\frac{P_{N}(t_{N-1})\rho_{N}-g_{N}(t_{N-1})}{P_{N}(t_{N-1})
+\mu_{N-1}e^{-\int_{t_{N-2}}^{t_{N-1}}\beta(t)\mathrm{d}t}}
\big[1-e^{-\int_{t_{N-2}}^{t_{N-1}}\beta(t)\mathrm{d}t}\big]\\
&+\displaystyle \frac{P_{N}(t_{N-1})+u_{N-1}}{P_{N}(t_{N-1})
+\mu_{N-1}e^{-\int_{t_{N-2}}^{t_{N-1}}\beta(t)\mathrm{d}t}}
\mathbb{E}[Y^{*}(t_{N-2})]
e^{\int_{t_{N-2}}^{t_{N-1}}[r(t)-\beta(t)]\mathrm{d}t}\bigg{)}.
\\
 \end{array}
\end{equation}
Combining equations (\ref{ee-2}) and (\ref{mean-4}), we have,
\begin{equation}\label{mu-2}
\begin{array}
[c]{rl}%
\mu_{N-1}=& \displaystyle \frac{ \big[1+P_{N}(t_{N-1})\rho_{N}-g_{N}(t_{N-1}) \big]
(e^{\int_{t_{N-2}}^{t_{N-1}}\beta(t)\mathrm{d}t}-1)  }{\mathbb{E}[Y^{*}(t_{N-1})]
-\mathbb{E}[Y^{*}(t_{N-2})]e^{\int_{t_{N-2}}^{t_{N-1}}r(t)\mathrm{d}t}}\\
&-\displaystyle \frac{\big[\mathbb{E}[Y^{*}(t_{N-1})]-\mathbb{E}[Y^{*}(t_{N-2})]
e^{\int_{t_{N-2}}^{t_{N-1}}[r(t)-\beta(t)]\mathrm{d}t}\big]P_{N}(t_{N-1})   e^{\int_{t_{N-2}}^{t_{N-1}}\beta(t)\mathrm{d}t}}{\mathbb{E}[Y^{*}(t_{N-1})]
-\mathbb{E}[Y^{*}(t_{N-2})]e^{\int_{t_{N-2}}^{t_{N-1}}r(t)\mathrm{d}t}}.
 \end{array}
\end{equation}
Again, based on constrained condition (\ref{mean-1}) of $\mathbb{E}[Y^*(t_{N-1})]=L_{N-1},\ \mathbb{E}[Y^*(t_{N-2})]=L_{N-2}$ and condition (\ref{con-1}), we can solve $\lambda^*_{N-1}$ and $\mu_{N-1}>0$.

Similar to the case $i=N-1$,  we can solve $\lambda^*_i,\mu_i, \ i=1,2,\cdots,N-1$ step by step from $N-1$ to $1$,
\begin{equation}\label{lam-3}
\begin{array}
[c]{rl}%
\lambda^*_{i}=& \displaystyle   \frac{P_{i+1}(t_{i})+\mu_{i}}{P_{i+1}(t_{i})
+\mu_{i}e^{-\int_{t_{i-1}}^{t_{i}}\beta(t)\mathrm{d}t}}\\
&\displaystyle +\mu_{i}\bigg{(}
\frac{P_{i+1}(t_{i})\rho_{i+1}-g_{i+1}(t_{i})}{P_{i+1}(t_{i})
+\mu_{i}e^{-\int_{t_{i-1}}^{t_{i}}\beta(t)\mathrm{d}t}}
\big[1-e^{-\int_{t_{i-1}}^{t_{i}}\beta(t)\mathrm{d}t}\big]\\
&+\displaystyle \frac{P_{i+1}(t_{i})+u_{i}}{P_{i+1}(t_{i})
+\mu_{i}e^{-\int_{t_{i-1}}^{t_{i}}\beta(t)\mathrm{d}t}}
\mathbb{E}[Y^{*}(t_{i-1})]
e^{\int_{t_{i-1}}^{t_{i}}[r(t)-\beta(t)]\mathrm{d}t}\bigg{)},
\\
 \end{array}
\end{equation}
and
\begin{equation}\label{mu-3}
\begin{array}
[c]{rl}%
\mu_{i}=& \displaystyle \frac{ \big[1+P_{i+1}(t_{i})\rho_{i+1}-g_{i+1}(t_{i}) \big]
(e^{\int_{t_{i-1}}^{t_{i}}\beta(t)\mathrm{d}t}-1)  }{\mathbb{E}[Y^{*}(t_{i})]
-\mathbb{E}[Y^{*}(t_{i-1})]e^{\int_{t_{i-1}}^{t_{i}}r(t)\mathrm{d}t}}\\
&-\displaystyle \frac{\big[\mathbb{E}[Y^{*}(t_{i})]-\mathbb{E}[Y^{*}(t_{i-1})]
e^{\int_{t_{i-1}}^{t_{i}}[r(t)-\beta(t)]\mathrm{d}t}\big]P_{i+1}(t_{i})e^{\int_{t_{i-1}}^{t_{i}}\beta(t)\mathrm{d}t}   }{\mathbb{E}[Y^{*}(t_{i})]
-\mathbb{E}[Y^{*}(t_{i-1})]e^{\int_{t_{i-1}}^{t_{i}}r(t)\mathrm{d}t}}.
 \end{array}
\end{equation}
Therefore, the optimal strategy $\pi^*(\cdot)$ of cost functional (\ref{cost-3-3}) is an optimal strategy of cost functional (\ref{cost-3-2}).  This completes the proof. $\ \  \  \   \   \  \   \  \Box$

\begin{remark}\label{re-1}
The conditions (\ref{con-1}) and (\ref{mean-3}) guarantee that  cost functional (\ref{cost-3-2}) has an optimal strategy with the parameters $\lambda^*$ and $\mu$. However, we haven't given the condition for $L_i,\ i=0,1,\cdots,N$ to guarantee $\mu_i>0,\ i=1,2,\cdots, N$ which satisfies conditions (\ref{con-1}). In the following section, we give the condition  for $L_1,L_2$ to guarantee $\mu_1,\mu_2>0$ for the case $N=2$ and solve $\lambda^*=(\lambda^*_1,\lambda^*_2), \ \mu=(\mu_1,\mu_2)$ by $\mathbb{E}[Y^{\pi}(t_1)]=L_1,\ \mathbb{E}[Y^{\pi}(t_2)]=L_2$.
\end{remark}

\section{Explicit solution and simulation}

In this section, we consider a simple example with $N=2$ which is used to verify the results in Theorem \ref{the-3} and investigate an explicit solution  for the parameters $\lambda^*=(\lambda_1,\lambda_2),\ \mu=(\mu_1,\mu_2),\ (\mathbb{E}[Y^{*}(t_1)],\mathbb{E}[Y^{*}(t_2)]),$ and variance $(\mathrm{Var}[Y^{*}(t_1)],\mathrm{Var}[Y^{*}(t_2)])$. Furthermore, we compare our multi-time state mean-variance model with classical mean-variance model.
\subsection{Explicit solution}
We suppose there are two assets, one bond and one stock, which are traded in the market.  Let $d=1,\ n=1,\ N=2$, the bond satisfies,
\begin{eqnarray*}
\left\{\begin{array}{rl}
\mathrm{d}R(t) & = r(t)R(t)\mathrm{d}t,\;\;  t>0,\\
 R(0) & = a_0>0,
\end{array}\right.
\end{eqnarray*}
and the stock asset is described by,
\begin{eqnarray*}
\left\{\begin{array}{rl} \mathrm{d}S(t) & \!\!\!=
b(t)S(t)\mathrm{d}t+ \displaystyle \sigma(t) S(t) \mathrm{d}W(t),\;\;
t>0,\\
 S(0) & \!\!\!= s_0>0.
\end{array}\right.
\end{eqnarray*}
Our target is to minimize  the following multi-time sate mean-variance problem:
\begin{equation}
\label{ss-cost-3-1}
J(\pi(\cdot))=\sum_{i=1}^2\bigg{(}\frac{\mu_i}{2} \mathrm{Var} (Y^{\pi}(t_i))-\mathbb{E}[Y^{\pi}(t_i)]\bigg{)},
\end{equation}
and a tractable auxiliary problem is given as follows:
 \begin{equation}
\label{ss-cost-3-2}
\hat{J}(\pi(\cdot))=\sum_{i=1}^2\mathbb{E}[\frac{\mu_i}{2} Y^{\pi}(t_i)^2-\lambda_i Y^{\pi}(t_i)].
\end{equation}

Based on the results in Theorem \ref{the-3} and formulas  (\ref{lam-1}) and (\ref{lam-3}), we set
\begin{equation*}
\begin{array}
[c]{rl}
\lambda^*_2=&\displaystyle {e^{\int_{t_{1}}^{t_2}\beta(t)\mathrm{d}t}}
+\mu_2\mathbb{E}[Y^{*}(t_{1})]e^{\int_{t_{1}}^{t_2}r(t)\mathrm{d}t};\\
\lambda^*_{1}=&\displaystyle\frac{P_{2}(t_{1})
+\mu_{1}}{P_{2}(t_{1})
+\mu_{1}e^{-\int_{0}^{t_{1}}\beta(t)\mathrm{d}t}}                \\
&+\displaystyle \mu_{1}\bigg{(}
\frac{P_{2}(t_{1})\rho_{2}-g_{2}(t_{1})}{P_{2}(t_{1})
+\mu_{1}e^{-\int_{0}^{t_{1}}\beta(t)\mathrm{d}t}}
\big[1-e^{-\int_{0}^{t_{1}}\beta(t)\mathrm{d}t}\big]\\
&+\displaystyle \frac{P_{2}(t_{1})+\mu_1}{P_{2}(t_{1})
+\mu_{1}e^{-\int_{0}^{t_{1}}\beta(t)\mathrm{d}t}} y
e^{\int_{0}^{t_{1}}[r(t)-\beta(t)]\mathrm{d}t}\bigg{)}.
 \end{array}
\end{equation*}
The  optimal strategy of model (\ref{ss-cost-3-2}) is given as follows:
\begin{equation*}
\begin{array}
[c]{rl}%
\pi^*(t)
=&\displaystyle \frac{b(t)-r(t)}{\sigma(t)^2}
\bigg[\big(\frac{\lambda_i^*}{\mu_i}-\frac{g_i(t_i)}{P_i(t_i)}\big)e^{-\int_t^{t_i}r(t)\mathrm{d}t}
-Y^{*}(t)\bigg], \ t_{i-1}<t\leq t_i,\ i=1,2,\\
 \end{array}
\end{equation*}
and
\begin{equation}\label{aux-1}
\begin{array}
[c]{rl}
&\mathbb{E}[Y^*(t_2)]=\displaystyle \frac{e^{\int_{t_{1}}^{t_2}\beta(t)\mathrm{d}t}-1}
{\mu_2}+{\mathbb{E}[Y^*(t_{1})]e^{\int_{t_{1}}^{t_2}r(t)\mathrm{d}t}};\\
&\mathbb{E}[Y^{*}(t_1)]=
ye^{\int_{0}^{t_1}[r(t)-\beta(t)]\mathrm{d}t}
+\displaystyle\big(\frac{\lambda_1^*}{\mu_1}-\frac
{P_{2}(t_1)(\rho_{1}-\rho_{2})+g_{2}(t_1)}{P_{2}(t_1)+\mu_1}\big)
\big(1-e^{-\int_{0}^{t_1}\beta(t)\mathrm{d}t}\big);\\
&\displaystyle \beta(t)=\bigg(\frac{b(t)-r(t)}{\sigma(t)}\bigg)^2,\ t\leq t_2,  \\
 \end{array}
\end{equation}
and $(P_2(t_1),g_2(t_1))$ satisfies the following Riccati equations,
\begin{equation}\label{ss-ric-eq-1}
\left\{\begin{array}{rl}
\!\!\!\mathrm{d}P_2(t)  & \!\!\!=\big[ \beta(t)-2r(t)\big]P_2(t) \mathrm{d}t,  \\
 \!\!\!P_2(t_2)& \!\!\!=\displaystyle \mu_2+P_{3}(t_2),\ t_{1}\leq t<t_2,
\end{array}\right.
\end{equation}
and related equations,
\begin{equation}\label{ss-ric-eq-2}
\left\{\begin{array}{rl}
\!\!\!\mathrm{d}g_2(t)  & =\big[ (\beta(t)-r(t))g_2(t)-\rho_2r(t)P_2(t)\big] \mathrm{d}t,  \\
 \!\!\!g_2(t_2)& =g_{3}(t_2)+P_{3}(t_2)(\rho_{2}-\rho_{3}),\ t_{1}\leq t< t_2,
\end{array}\right.
\end{equation}
where $P_{3}(t_2)=0,\ g_{3}(t_2)=0,\ \rho_{3}=0$. By a simple calculation, we can obtain that
\begin{equation}\label{para-1}
\begin{array}
[c]{rl}
&P_2(t_1)=\mu_2e^{\int_{t_1}^{t_2}[2r(t)-\beta(t)]\mathrm{d}t};\\
&\displaystyle \frac{\lambda_2^*}{\mu_2}-\frac{g_2(t_2)}{P_2(t_2)}=\frac{\lambda_2^*}{\mu_2};  \\
&\displaystyle \frac{\lambda_1^*}{\mu_1}-\frac{g_1(t_1)}{P_1(t_1)}=\frac
{\lambda_1^*+\lambda^*_2e^{\int_{t_1}^{t_2}[r(t)-\beta(t)]\mathrm{d}t}}
{\mu_1+\mu_2e^{\int_{t_1}^{t_2}[2r(t)-\beta(t)]\mathrm{d}t}}.  \\
 \end{array}
\end{equation}
Combining formulas (\ref{aux-1})  and (\ref{para-1}), it follows that
\begin{equation}\label{aux-2}
\begin{array}
[c]{rl}
\mathbb{E}[Y^{*}(t_1)]=&
ye^{\int_{0}^{t_1}[r(t)-\beta(t)]\mathrm{d}t}
+\displaystyle\frac
{\lambda_1^*+\lambda^*_2e^{\int_{t_1}^{t_2}[r(t)-\beta(t)]\mathrm{d}t}}
{\mu_1+\mu_2e^{\int_{t_1}^{t_2}[2r(t)-\beta(t)]\mathrm{d}t}}
\big[1-e^{-\int_{0}^{t_1}\beta(t)\mathrm{d}t}\big];\\
\mathbb{E}[Y^*(t_2)]=&\displaystyle \frac{e^{\int_{t_{1}}^{t_2}\beta(t)\mathrm{d}t}-1}
{\mu_2}+{\mathbb{E}[Y^*(t_{1})]e^{\int_{t_{1}}^{t_2}r(t)\mathrm{d}t}};\\
\lambda^*_2=&\displaystyle {e^{\int_{t_{1}}^{t_2}\beta(t)\mathrm{d}t}}
+\mu_2\mathbb{E}[Y^{*}(t_{1})]e^{\int_{t_{1}}^{t_2}r(t)\mathrm{d}t};\\
\lambda^*_{1}=&\displaystyle\frac{P_{2}(t_{1})
+\mu_{1}}{P_{2}(t_{1})
+\mu_{1}e^{-\int_{0}^{t_{1}}\beta(t)\mathrm{d}t}}                \\
&+\displaystyle \mu_{1}\bigg{(}
\frac{P_{2}(t_{1})\rho_{2}-g_{2}(t_{1})}{P_{2}(t_{1})
+\mu_{1}e^{-\int_{0}^{t_{1}}\beta(t)\mathrm{d}t}}
\big[1-e^{-\int_{0}^{t_{1}}\beta(t)\mathrm{d}t}\big]\\
&+\displaystyle \frac{P_{2}(t_{1})+\mu_1}{P_{2}(t_{1})
+\mu_{1}e^{-\int_{0}^{t_{1}}\beta(t)\mathrm{d}t}} y
e^{\int_{0}^{t_{1}}[r(t)-\beta(t)]\mathrm{d}t}\bigg{)}.
 \end{array}
\end{equation}

In the following, we set $T=2,\ y=1$, $t_1=1$, and $t_2=2$. Let $r(t)=r,\ b(t)=b,\ \sigma(t)=\sigma,\ \beta(t)=\beta$, where $0\leq t \leq T$. From formulas (\ref{aux-2}), we have
\begin{equation}\label{aux-3}
\begin{array}
[c]{rl}
\mathbb{E}[Y^{*}(1)]=&
e^{r-\beta}
+\displaystyle\frac
{\lambda_1^*+\lambda^*_2e^{r-\beta}}
{\mu_1+\mu_2e^{2r-\beta}}
\big(1-e^{-\beta}\big);\\
\mathbb{E}[Y^*(t_2)]=&\displaystyle \frac{e^{\beta}-1}
{\mu_2}+\mathbb{E}[Y^*(t_{1})]e^{r};\\
\lambda^*_2=& \displaystyle e^{\beta}
+\mu_2\mathbb{E}[Y^{*}(1)]e^{r};\\
\lambda^*_{1}=&\displaystyle \mu_{1}\bigg{(}\displaystyle\frac{\lambda^*_2(e^r-e^{r-\beta})}{\mu_2e^{2r}+\mu_1}
+\displaystyle\frac{\mu_2e^{3r-\beta}+\mu_1e^r}{\mu_2e^{2r}+\mu_1}\bigg{)}\\
&+\displaystyle\frac{\mu_2e^{2r}+\mu_1e^{\beta}}{\mu_2e^{2r}+\mu_1}.
 \end{array}
\end{equation}

\begin{remark}
Let $\mathbb{E}[Y^{*}(1)]=L_1,\mathbb{E}[Y^{*}(2)]=L_2$, and
\begin{equation}\label{ass-1}
\begin{array}
[c]{ll}
L_2>L_1e^r>e^{2r};\\
(L_2-L_1e^r)e^{\beta}>(L_1-e^r)e^r.
 \end{array}
\end{equation}
Note that,  the condition $L_2>L_1e^r>e^{2r}$ guarantees that the constraints on the mean values $\mathbb{E}[Y^{*}(1)]=L_1,\mathbb{E}[Y^{*}(2)]=L_2$ are bigger than the return which is invested into the risk-free asset bond, while the condition $(L_2-L_1e^r)e^{\beta}>(L_1-e^r)e^r$ guarantees the parameter $\mu_1>0$ in technique.

\end{remark}

Applying formulas (\ref{aux-3}), by a simple calculation, one obtains
\begin{equation}\label{aux-4}
\begin{array}
[c]{rl}
\mu_2=&\displaystyle \frac{e^{\beta}-1}{L_2-L_1e^r};\\
\lambda^*_2=&
\displaystyle \frac{L_2e^{\beta}-L_1e^{r}}{L_2-L_1e^r};\\
\mu_1=&\displaystyle \frac{(e^{\beta}-1)(e^{\beta}+\lambda_2^*e^r)-(L_1e^{\beta}-e^r)e^{2r}\mu_2}{(L_1-e^r)e^{\beta}};\\
\lambda^*_{1}=&\displaystyle \mu_{1}\bigg{(}\displaystyle\frac{\lambda^*_2(e^r-e^{r-\beta})}{\mu_2e^{2r}+\mu_1}
+\displaystyle\frac{\mu_2e^{3r-\beta}+\mu_1e^r}{\mu_2e^{2r}+\mu_1}\bigg{)}\\
&+\displaystyle\frac{\mu_2e^{2r}+\mu_1e^{\beta}}{\mu_2e^{2r}+\mu_1}.
 \end{array}
\end{equation}
Based on Theorem \ref{the-3}, applying the formula (\ref{para-1}), we can obtain  the related optimal strategy for the multi-time state mean-variance model (\ref{ss-cost-3-1}) with the constraints on means $\mathbb{E}[Y^{*}(1)]=L_1,\ \mathbb{E}[Y^{*}(2)]=L_2$,
\begin{equation}\label{opti-1}
\pi^*(t)=\left\{\begin{array}{ll}
\displaystyle \frac{b-r}{\sigma^2}
\bigg[\frac
{\lambda_1^*+\lambda^*_2e^{r-\beta}}
{\mu_1+\mu_2e^{2r-\beta}}e^{r(t-{1})}
-Y^{*}(t)\bigg],\ {0}\leq t\leq 1;\\
\displaystyle \frac{b-r}{\sigma^2}
\bigg[\frac{\lambda_2^*}{\mu_2}e^{r(t-{2})}
-Y^{*}(t)\bigg],\ {1}< t\leq  2.\\
\end{array}\right.
\end{equation}
Thus,  $\mathbb{E}[Y^{*}(\cdot)]$  satisfies
\begin{equation*}
\mathbb{E}[{Y}^{*}(t)]=\left\{\begin{array}
[c]{rl}%
&   \displaystyle e^{(r-\beta)t}
+\frac
{\lambda_1^*+\lambda^*_2e^{r-\beta}}
{\mu_1+\mu_2e^{2r-\beta}}[e^{r(t-1)}-e^{(r-\beta)t-r}],\ 0\leq t\leq 1; \\
& \mathbb{E}[{Y}^{*}(1)]e^{(r-\beta)(t-1)}
\displaystyle+\frac{\lambda_2^*}{\mu_2}[e^{r(t-2)}-e^{r(t-2)-\beta(t-1)}],\ 1< t\leq 2,\\
\end{array}\right.
\end{equation*}
and from Lemma \ref{lem-1}, the variance of $Y^{*}(\cdot)$ at $t_1=1,\ t_2=2$ are given as follows:
\begin{equation}\label{ss-2-1}
\begin{array}
[c]{rl}%
&\mathrm{Var}(Y^{*}(1))=\displaystyle\frac{\bigg(\mathbb{E}[Y^{*}(1)]
-e^{r}\bigg)^2}
{e^{\beta}-1};\\
&\mathrm{Var}(Y^{*}(2))=\mathrm{Var}(Y^{*}(1))
e^{2r-\beta}+\displaystyle\frac{\bigg(\mathbb{E}[Y^{*}(2)]
-\mathbb{E}[Y^{*}(1)]e^{r}\bigg)^2}
{e^{\beta}-1}.\\
 \end{array}
\end{equation}

\bigskip

Now, we show the results of  case $N=1$ which is  the classical mean-variance model:
\begin{equation*}
\begin{array}
[c]{rl}
&\mathbb{E}[{Y}^{\#}(2)]=L_2;  \\
&\mu=\displaystyle \frac{e^{2\beta}-1}{L_2-e^{2r}};\\
&\lambda^*=\displaystyle \frac{L_2e^{2\beta}-e^{2r}}{L_2-e^{2r}},\\
 \end{array}
\end{equation*}
where the related optimal strategy is
\begin{equation}
\pi^{\#}(t)=\displaystyle \frac{b-r}{\sigma^2}
\bigg[\frac{\lambda^*}{\mu}e^{r(t-{2})}
-Y^{\#}(t)\bigg],\ 0\leq  t\leq  2.
\end{equation}
The mean $\mathbb{E}[Y^{\#}(\cdot)]$ and variance $\mathrm{Var}(Y^{\#}(\cdot))$  satisfy
\begin{equation}\label{ss-2-2}
\begin{array}
[c]{rl}%
&   \displaystyle \mathbb{E}[{Y}^{\#}(t)]=e^{(r-\beta)t}
+\frac{\lambda^*e^{r(t-2)}}{\mu}[1-e^{-\beta t}];\\
&\mathrm{Var}(Y^{\#}(t))=\displaystyle\frac{\big(\mathbb{E}[Y^{\#}(t)]
-e^{rt}\big)^2}
{e^{\beta t}-1},\ 0\leq t\leq 2,\\
\end{array}
\end{equation}
where $\mathbb{E}[{Y}^{\#}(2)]=L_2$. Based on formulas (\ref{ss-2-1}) and (\ref{ss-2-2}),  we have the following comparison results for  $(\mathrm{Var}(Y^{*}(1)),\mathrm{Var}(Y^{*}(2)))$ and $(\mathrm{Var}(Y^{\#}(1)),\mathrm{Var}(Y^{\#}(2)))$:
\begin{corollary}\label{cor-1}
Suppose $L_1$ and $L_2$   satisfy condition (\ref{ass-1}), one obtains
\begin{equation*}
\begin{array}
[c]{rl}%
&\mathrm{Var}(Y^{*}(1))<\mathrm{Var}(Y^{\#}(1));\\
&\mathrm{Var}(Y^{*}(2))>\mathrm{Var}(Y^{\#}(2)).\\
\end{array}
\end{equation*}
Furthermore, if
$$
\displaystyle \frac{L_2+e^{2r-\beta}+1}{e^{r-\beta}+e^r+e^{-r}}\leq L_1< \frac{L_2+e^{2r-\beta}}{e^{r-\beta}+e^r},
$$
we have
$$
\mathrm{Var}(Y^{*}(1))+\mathrm{Var}(Y^{*}(2))<\mathrm{Var}(Y^{\#}(1))+\mathrm{Var}(Y^{\#}(2)).
$$
\end{corollary}
\noindent \textbf{Proof}: By equality  (\ref{ss-2-2}), we have
$$
\mathbb{E}[{Y}^{\#}(1)]=\frac{L_2e^{\beta-r}+e^r}{e^{\beta}+1}.
$$
Applying the condition $(L_2-L_1e^r)e^{\beta}>(L_1-e^r)e^r$ in (\ref{ass-1}), one obtains
$$
e^r<\mathbb{E}[{Y}^{*}(1)]=L_1<\frac{L_2e^{\beta-r}+e^r}{e^{\beta}+1}=\mathbb{E}[{Y}^{\#}(1)],
$$
it follows that,
$$
\mathrm{Var}(Y^{*}(1))<\mathrm{Var}(Y^{\#}(1)).
$$

By formula (\ref{ss-2-1}), we have
\begin{equation*}
\begin{array}
[c]{rl}%
\mathrm{Var}(Y^{*}(2))=&\displaystyle\frac{\big(L_1-e^r\big)^2}{e^{\beta}-1}
e^{2r-\beta}+\frac{\big(L_2-L_1e^{r}\big)^2}{e^{\beta}-1}\\
=&
\displaystyle\frac{  [e^{2r-\beta}+e^{2r}]L_1^2-2e^r[L_2+e^{2r-\beta}]L_1             +L_2^2+e^{4r-\beta}}{e^{\beta}-1}\\
=& \displaystyle \frac{e^{2r-\beta}+e^{2r}}{e^{\beta}-1} \bigg[L_1-  \frac{L_2e^{\beta-r}+e^r}{e^{\beta}+1} \bigg]^2+  \frac{\big(L_2-e^{2r} \big)^2}{e^{2\beta}-1}.                             \\
\end{array}
\end{equation*}
From equality (\ref{ss-2-2}), one obtains
$$
\mathrm{Var}(Y^{\#}(2)) =\frac{\big(L_2-e^{2r} \big)^2}{e^{2\beta}-1}.
$$
It follows that,
$$
\mathrm{Var}(Y^{*}(2))>\mathrm{Var}(Y^{\#}(2)).
$$

Furthermore, we have
\begin{equation*}
\begin{array}
[c]{rl}%
&\mathrm{Var}(Y^{*}(1))+\mathrm{Var}(Y^{*}(2))\\
=&\displaystyle\frac{\big(L_1-e^r\big)^2}{e^{\beta}-1}
[e^{2r-\beta}+1]+\frac{\big(L_2-L_1e^{r}\big)^2}{e^{\beta}-1}\\
=&
\displaystyle\frac{  [e^{2r-\beta}+e^{2r}+1]L_1^2-2e^r[L_2+e^{2r-\beta}+1]L_1             +L_2^2+e^{4r-\beta}+e^{2r}}{e^{\beta}-1}.\\
\end{array}
\end{equation*}
It follows that $\mathrm{Var}(Y^{*}(1))+\mathrm{Var}(Y^{*}(2))$ admits the minimum values at
$$
L_1=\frac{L_2+e^{2r-\beta}+1}{e^{r-\beta}+e^r+e^{-r}},
$$
Again, applying condition (\ref{ass-1}), we have
$$
\frac{L_2+e^{2r-\beta}+1}{e^{r-\beta}+e^r+e^{-r}}<\frac{L_2+e^{2r-\beta}}{e^{r-\beta}+e^r}=\mathbb{E}[{Y}^{\#}(1)].
$$
Notice that, if
$$
L_1=\frac{L_2+e^{2r-\beta}}{e^{r-\beta}+e^r},
$$
one obtains,
$$
\mathrm{Var}(Y^{*}(1))+\mathrm{Var}(Y^{*}(2))=\mathrm{Var}(Y^{\#}(1))+\mathrm{Var}(Y^{\#}(2)).
$$
Thus if
$$
\displaystyle \frac{L_2+e^{2r-\beta}+1}{e^{r-\beta}+e^r+e^{-r}}\leq L_1< \frac{L_2+e^{2r-\beta}}{e^{r-\beta}+e^r},
$$
we have
$$
\mathrm{Var}(Y^{*}(1))+\mathrm{Var}(Y^{*}(2))<\mathrm{Var}(Y^{\#}(1))+\mathrm{Var}(Y^{\#}(2)).
$$
\noindent This completes the proof. $\ \ \ \ \ \ \ \Box$

\bigskip

\subsection{Simulation analysis}

 Let $r=0.04,\ b=0.12,\ \sigma=0.2,\ \beta=0.16$, we show the simulation results of the case $N=2$, and case $N=1$, where  case $N=1$ is same with the classical continuous time mean-variance model.

\begin{figure}[H]
 \caption{Comparing  the  values of $\mathbb{E}[Y^*(\cdot)]$ and $\mathbb{E}[Y^{\#}(\cdot)]$}
 \label{fig1}
\begin{center}
\includegraphics[width=4.5 in]{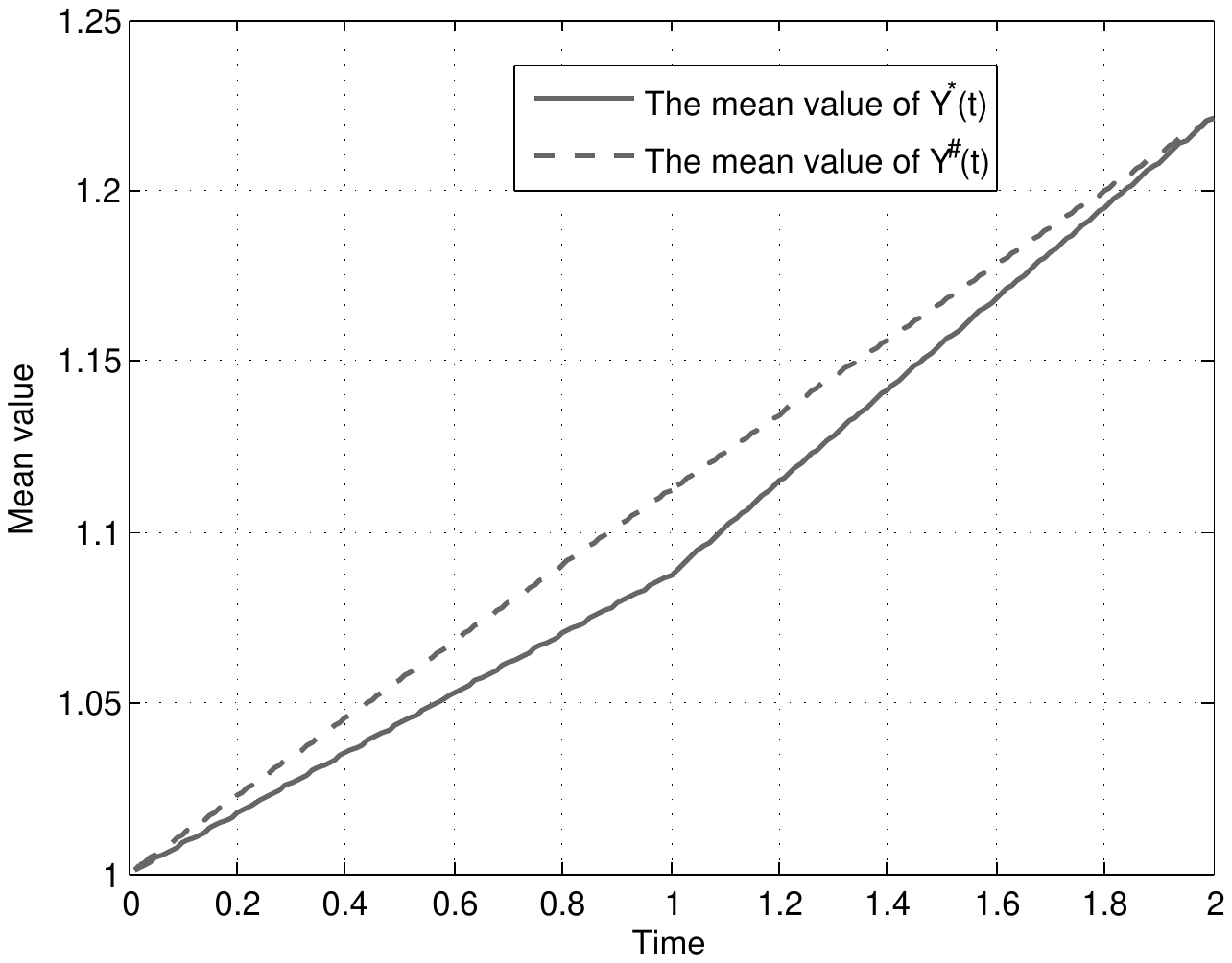}
\end{center}
\end{figure}

In Figure \ref{fig1}, we take $L_2=e^{5r},\ L_1=e^{2.1r}$ which satisfies conditions  (\ref{ass-1}). The expectations of $Y^*(\cdot)$ and  $Y^{\#}(\cdot)$ are given as follows, respectively,
\begin{equation*}
 \mathbb{E}[{Y}^{*}(t)]=\left\{\begin{array}
[c]{rl}%
&   \displaystyle e^{(r-\beta)t}
+\frac
{\lambda_1^*+\lambda^*_2e^{r-\beta}}
{\mu_1+\mu_2e^{2r-\beta}}[e^{r(t-1)}-e^{(r-\beta)t-r}],\ 0\leq t\leq 1; \\
& \mathbb{E}[{Y}^{*}(1)]e^{(r-\beta)(t-1)}
\displaystyle+\frac{\lambda_2^*}{\mu_2}[e^{r(t-2)}-e^{r(t-2)-\beta(t-1)}],\ 1< t\leq 2,\\
\end{array}\right.
\end{equation*}
and
$$
\displaystyle \mathbb{E}[{Y}^{\#}(t)]=e^{(r-\beta)t}
+\frac{\lambda^*e^{r(t-2)}}{\mu}[1-e^{-\beta t}], \   0\leq  t\leq 2.
$$
From  conditions  (\ref{ass-1}), we obtain  $\mathbb{E}[{Y}^{*}(1)]=L_1< \mathbb{E}[{Y}^{\#}(1)]$ and thus,
$$
\mathbb{E}[{Y}^{*}(t)]<\mathbb{E}[{Y}^{\#}(t)],\ 0< t< 2.
$$
These results shows that  if we want to minimize the variances of the wealth  at times $t_1=1,\ t_2=2$ together, the means of the investment portfolio may be smaller than that of classical mean-variance model in continuous time.

\begin{figure}[H]
 \caption{Comparing the values $Y^*(\cdot)$ and $Y^{\#}(\cdot)$}
 \label{fig2}
\begin{center}
\includegraphics[width=2.65 in]{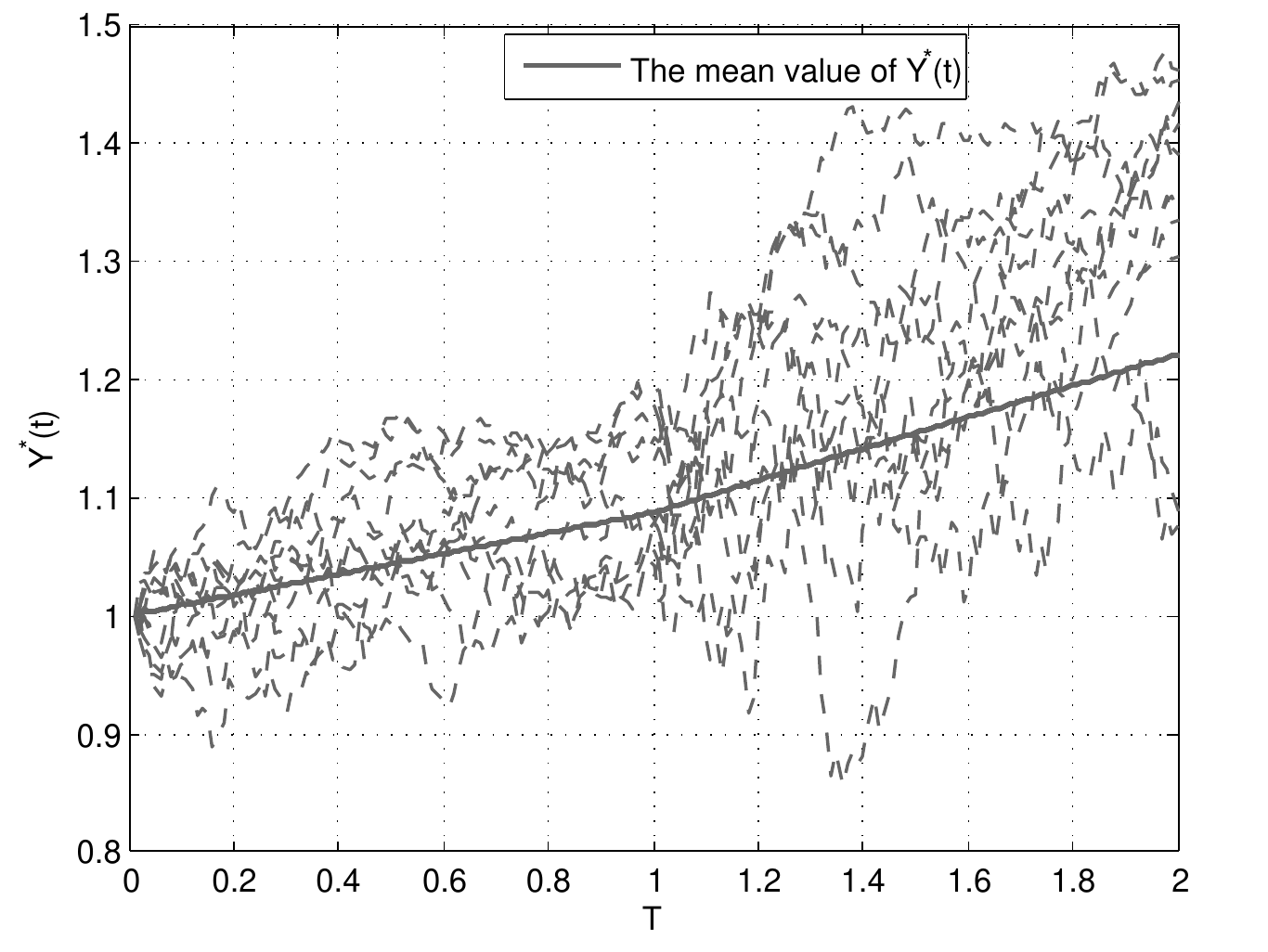}
\includegraphics[width=2.65 in]{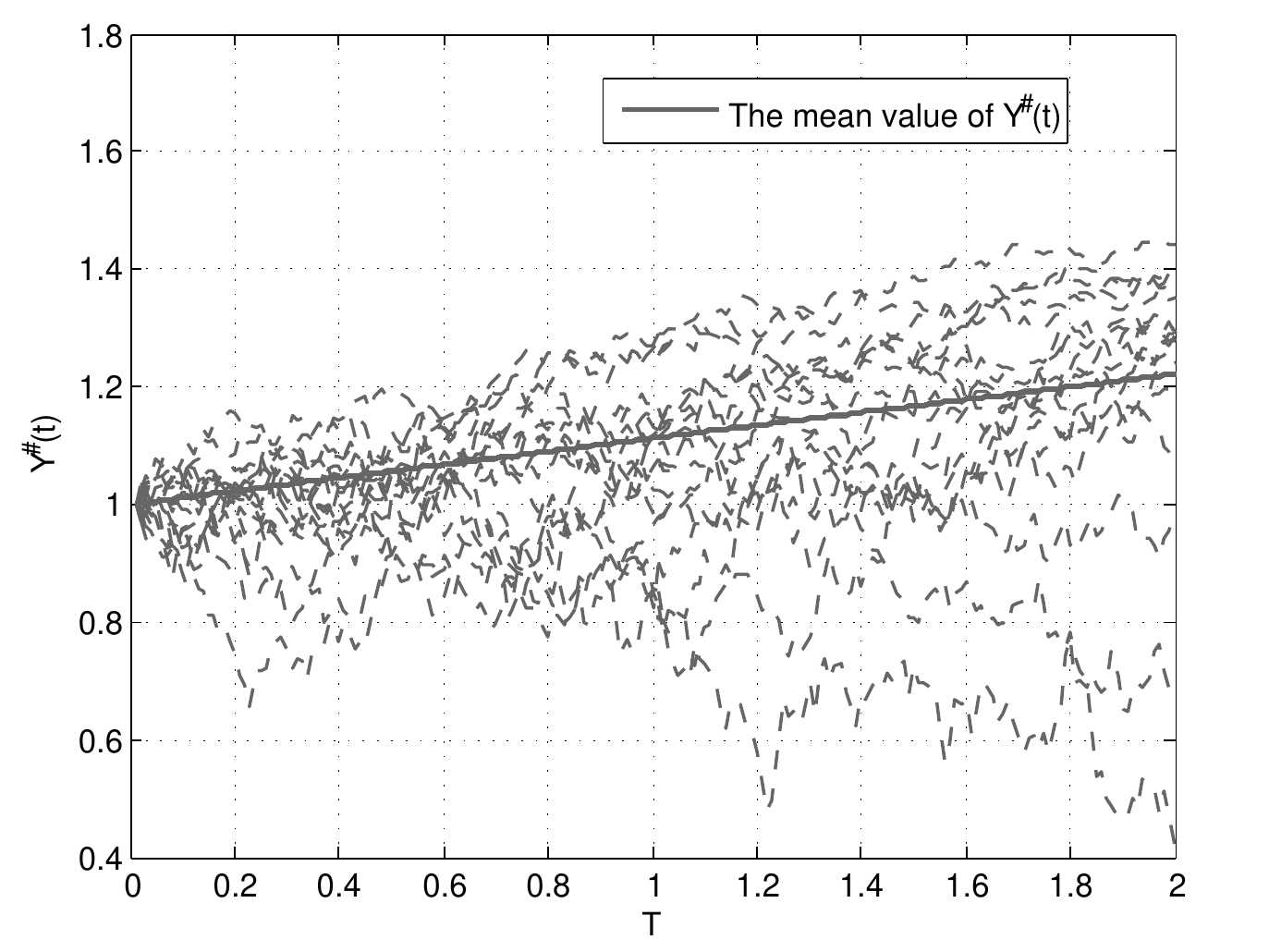}
\end{center}
\end{figure}
{In Figure \ref{fig2}, we plot  the values  $Y^*(\cdot)$ and $Y^{\#}(\cdot)$ in pathwise. The left one shows that the pathwise of the function $Y^*(\cdot)$ along with  $\mathbb{E}[Y^*(\cdot)]$, while the right one shows that of $Y^{\#}(\cdot)$}. We can see that the variance of $Y^*(1)$ is bigger than that of $Y^*(1)$, and the variance of $Y^*(2)$ is almost the same as that of $Y^*(2)$. These phenomena  verify the results of  Corollary   \ref{cor-1}. In addition,  In Figure \ref{fig1}, we can see that $\mathbb{E}[{Y}^{*}(t)]<\mathbb{E}[{Y}^{\#}(t)],\ 0<t\leq 1$, while Figure \ref{fig2} shows that the variance of $Y^*(\cdot)$ is smaller than that of $Y^{\#}(\cdot)$ before time $1$.

\begin{figure}[H]
 \caption{The average of Maximum-Drawdown of $Y^*(\cdot)$ along with $\theta$}
 \label{fig3}
\begin{center}
\includegraphics[width=4.6 in]{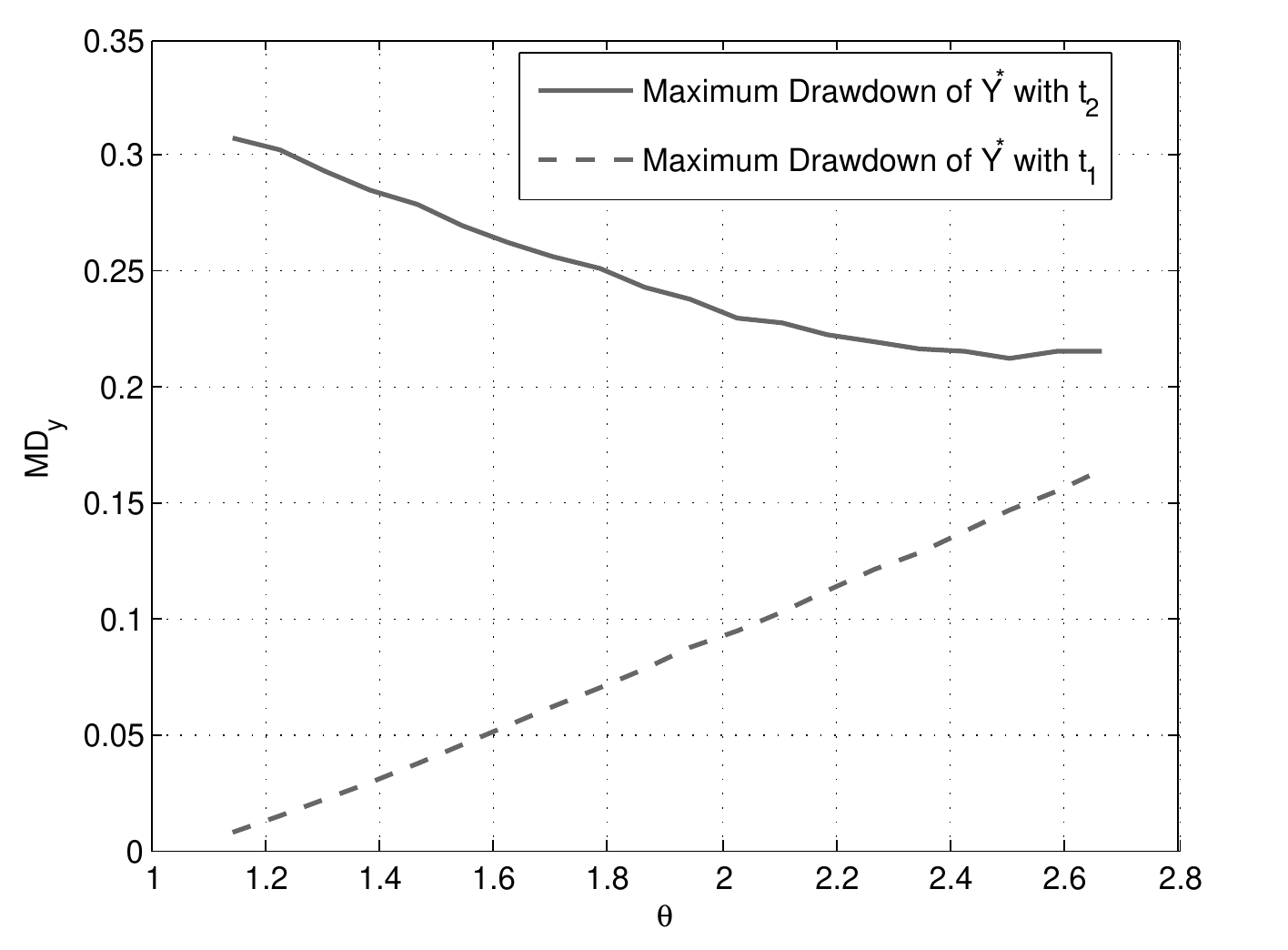}
\end{center}
\end{figure}
 In Figure \ref{fig3}, we plot the function of $\mathbb{E}[\mathrm{MD}_{Y^{*}}]$ along with $\theta\in [1.145, 2.665]$, where
$$
\mathrm{MD}^h_{Y^{*}}=\mathrm{esssup}\ \{ z  \ | \  z=Y^{*}(t)- Y^{*}(s),\ 0\leq t\leq s\leq h  \},
$$
$0<h\leq 2$, and
$$
L_1=e^{\theta r},\quad  \mathbb{E}[Y^{\#}(1)]=e^{2.665  r}.
$$
We can see that $\mathbb{E}[\mathrm{MD}^{t_2}_{Y^{*}}]$ is decreasing with $\theta\in [1.145, 2.505]$, increasing with  $\theta\in [2.505, 2.665]$  and thus decreasing with $L_1\in [e^{1.145  r},e^{2.505  r}]$, increasing with $L_1\in [e^{2.505  r},e^{2.665  r}]$, while  $\mathbb{E}[\mathrm{MD}^{t_1}_{Y^{*}}]$ is increasing with $L_1\in [e^{1.145  r},e^{2.665  r}]$, where $t_1=1,\ t_2=2$.

\section{Conclusion}
For given $0=t_0<t_1<\cdots<t_N=T$, to reduce the variance of the mean-variance model at the multi-time state $(Y^{\pi}(t_1),\cdots,Y^{\pi}(t_N))$, we propose a multi-time state mean-variance model with a constraint on  the multi-time state mean value. In the proposed model, we solve the multi-time state mean-variance model by introducing a new sequence of Riccati equations.

Our main results are  as follows:
\begin{itemize}
\item  We can use the multi-time state mean-variance model to manage the risk of the investment portfolio along the multi-time $0=t_0<t_1<\cdots<t_N=T$.

 \item  A new sequence of Riccati equations which are connected by a jump boundary condition are introduced, based on which we find an optimal strategy for  the multi-time state mean-variance model.

 \item Furthermore, the relationship of the means and variances of this multi-time state mean-variance model is established and is similar to the classical mean-variance model.

\item An example is employed to show that  minimizing  the variances for  multi-time state can affect the average value of  Maximum-Drawdown of the investment portfolio.

\end{itemize}

\bibliography{var}

\end{document}